# Necessary and sufficient conditions for Boolean satisfiability

Stepan G. Margaryan

*e-mail: stmargaryan@gmail.com*
*Yerevan, Armenia*

**Abstract.**

This is the second in a series of articles aimed at exploring the relationship between the complexity classes of P and NP.

The research in this article aims to find conditions of an algorithmic nature that are necessary and sufficient to transform any Boolean function in conjunctive normal form into a specific form that guarantees the satisfiability of this function.

To find such conditions, we use the concept of a special covering of a set introduced in [13], and investigate the connection between this concept and the notion of satisfiability of Boolean functions.

As shown, the problem of existence of a special covering for a set is equivalent to the Boolean satisfiability problem. Thus, an important result is the proof of the existence of necessary and sufficient conditions that make it possible to find out if there is a special covering for the set under the special decomposition. This result allows us to formulate the necessary and sufficient algorithmic conditions for Boolean satisfiability, considering the function in conjunctive normal form as a set of clauses.

In parallel, as a result of the aforementioned algorithmic procedure, we obtain the values of the variables that ensure the satisfiability of this function.

The terminology used related to graph theory, set theory, Boolean functions and complexity theory is consistent with the terminology in [1], [2], [3], [4]. The newly introduced terms are not found in use by other authors and do not contradict to other terms.

## 1. Introduction

Despite the fact that the relationship between the complexity classes $P$ and $NP$ is still an open problem in complexity theory, the problem of decidability of a propositional formula in conjunctive normal form has been extensively studied both from the point of view of theoretical research and from the point of view of improving specific algorithms, included probabilistic algorithms.

In particular, various complexity classes based on time and space complexity, as well as the relations between them, are widely studied in [1]. In [7], algorithms with different approaches and methods are described and upper bounds for the worst cases are given.

A well-known theoretical study is Schaefer's result in [11], where specific classes of Boolean formulas with certain restrictions are considered. If the formula belongs to one of these classes, then it is decidable in polynomial time. In fact, Schaefer's dichotomy theorem shows that, depending on the restrictions in a propositional formula, the problem either belongs to the class $P$ or is $NP$-complete. Later, in [6], Schaefer's result was refined and extended considering $AC^0$ isomorphisms, and it was shown that with this approach there can be more different complexity classes.

However, some problems arise when recognizing that a formula belongs to a given class. It is not always "easy" to check if a given formula belongs to a certain class. The complexity of the recognition problem depends on how the constraints are specified [5]. In [8], recognition criteria related to this problem are also formulated.

The research in this article is aimed at finding necessary and sufficient conditions for transforming a Boolean function in conjunctive normal form into a specific form that guarantees the satisfiability of this function.

We will prove the existence of such conditions that are of an algorithmic nature. To prove this, we use the concept of a special decomposition of a set and the concept of a special covering for a set, introduced in [13], and study the connection of these concepts with the satisfiability of Boolean functions in conjunctive normal form.

The problem of existence of a special covering of a set is equivalent to the Boolean satisfiability problem [13]. Thus, an important result is the proof of the existence of necessary and sufficient conditions that make it possible to find out if there exists a special covering for the set under the special decomposition. This result allows us to formulate the necessary and sufficient conditions for Boolean satisfiability, considering the function in conjunctive normal form as a set of clauses.

In addition, we will also describe a simple procedure for finding the maximum number of satisfiable clauses in a function if the function is not satisfiable.

To formulate the same result in term of Boolean function we introduce the concept of proportional conjunctive normal form of a function, which is a conjunctive normal form of a function with the condition that each clause contains a negative literal or each clause contains a positive literal.

Thus, we obtain that a Boolean function represented in conjunctive normal form is satisfiable if and only if it is transformed into a function in proportional conjunctive normal form by literal inversion.

In parallel, if the function is satisfiable, then as a result of the procedure for representing the function in proportional-conjunctive normal form, the values of the variables are obtained that ensure the satisfiability of this function.

Also, if the function is not satisfiable, then the general procedure comes to this conclusion.

## 2. Basic Definitions and Denotes

For convenience, we recall some basic definitions and denotes from [13].

$$S = \{e_1, e_2, \ldots, e_m\}$$

is a nonempty set of $m$ elements for some natural number $m$.

We assume that for some natural number $n$, $n$ arbitrary ordered pairs of arbitrary subsets of the set $S$ are given. For some $\alpha \in \{0,1\}$ we will use the notation $(M_i^{\alpha}, M_i^{1-\alpha})$ for these ordered pairs, where $i \in \{1, \ldots, n\}$, and the superscript 1- $\alpha$ means 1 when $\alpha$ = 0, and 0, when $\alpha$ = 1. The subset $M_i^{\alpha}$ will be called the $\alpha$-component of the ordered pair $(M_i^{\alpha}, M_i^{1-\alpha})$ for $\alpha \in \{0,1\}$.

Denote by $d_nS$ the ordered set of those ordered pairs:

$$d_nS = \{(M_1^{\alpha}, M_1^{1-\alpha}), (M_2^{\alpha}, M_2^{1-\alpha}), \ldots, (M_n^{\alpha}, M_n^{1-\alpha})\},$$

*Definition* 2.1. The set $d_nS$ will be called a special decomposition of the set $S$, if

(2.1.1) $\forall\, i \in \{1, \ldots, n\}$ $(M_i^{\alpha} \cap M_i^{1-\alpha}) = \emptyset$,

(2.1.2) $\forall\, i \in \{1, \ldots, n\}$ $(M_i^{\alpha} \neq \emptyset)$ or $M_i^{1-\alpha} \neq \emptyset)$,

(2.1.3) $\bigcup_{i=1}^{n}(M_i^{\alpha} \cup M_i^{1-\alpha}) = S$.

*Definition* 2.2. Let the set $d_nS$ be a special decomposition of the set $S$.

For some $\alpha_1, \alpha_2, \ldots, \alpha_n$, where $\alpha_i \in \{0,1\}$, the ordered set

$$c_nS = \{M_1^{\alpha_1}, M_2^{\alpha_2}, \ldots, M_n^{\alpha_n}\},$$

will be called a special covering for the set $S$ under the special decomposition $d_nS$, if

$$\bigcup_{i=1}^{n} M_i^{\alpha_i} = S.$$

For an ordered set $d_nS$ and for any $\alpha \in \{0, 1\}$ we denote:

$M^{\alpha} = \bigcup_{i=1}^{n} M_i^{\alpha}$.

$sM^{\alpha} = \{M_1^{\alpha}, \ldots, M_n^{\alpha}\}$. We call $sM^{\alpha}$ the ordered set of $\alpha$-components of the decomposition $d_nS$ or, briefly, the $\alpha$-domain of the decomposition $d_nS$.

We say that an element $e \in S$ is in the $\alpha$-domain of a special decomposition if $e$ occurs in some $\alpha$-components of this decomposition.

$(i_1, \ldots, i_k)I(d_nS)$ is the ordered set obtained by permuting the components of ordered pairs $(M_{i_1}^{\alpha}, M_{i_1}^{1-\alpha}), \ldots, (M_{i_k}^{\alpha}, M_{i_k}^{1-\alpha})$ of the set $d_nS$. We will call it the $I$-transformation of the decomposition $d_n$S using the ordered pairs $(M_{i_1}^{\alpha}, M_{i_1}^{1-\alpha}), \ldots, (M_{i_k}^{\alpha}, M_{i_k}^{1-\alpha})$.

$(i_1, \ldots, i_k)$s$M^{\alpha}$ is the ordered set obtained by replacing the subsets $M_{i_1}^{\alpha}, \ldots, M_{i_k}^{\alpha}$ of the set $sM^{\alpha}$ by the subsets $M_{i_1}^{1-\alpha}, \ldots, M_{i_k}^{1-\alpha}$, respectively. We will call $(i_1, \ldots, i_k)$s$M^{\alpha}$ the set of $\alpha$-components or, briefly, $\alpha$-domain of the ordered set $(i_1, \ldots, i_k)I(d_nS)$.

$(i_1, \ldots, i_k)M^{\alpha}$ is the set of elements included in the $\alpha$-components of the ordered set $(i_1, \ldots, i_k)I(d_nS)$. We call $(i_1, \ldots, i_k)M^{\alpha}$ the set of elements included in $\alpha$-domain of the ordered set $(i_1, \ldots, i_k)I(d_nS)$.

*If $\alpha$-domain of the corresponding special decomposition is a special covering for the set $S$, then we will call it an $M^{\alpha}$-covering for the set $S$.*

## 3. Replaceability of Subsets

We assume that some nonempty set $S = \{e_1, e_2, \ldots, e_m\}$ of $m$ elements is given and, in addition,

$$M_1^{\alpha},\ M_1^{1-\alpha}, \ldots, M_i^{\alpha},\ M_i^{1-\alpha}, \ldots, M_n^{\alpha},\ M_n^{1-\alpha}$$

are arbitrary subsets of the set $S$ such that the set

$$d_nS = \{(M_1^{\alpha},\ M_1^{1-\alpha}), \ldots, (M_i^{\alpha}, M_i^{1-\alpha}), \ldots, (M_n^{\alpha},\ M_n^{1-\alpha})\}$$

is a special decomposition of the set $S$. The subsets forming a special decomposition of the set $S$ will also be called subsets of the given decomposition.

We will search for special covering for the set $S$, under the special decomposition $d_nS$, if none of the sets $sM^{\alpha}$ and $sM^{1-\alpha}$ covers the set $S$.

According to Lemma 1.8 in [13], there exists a special covering for the set $S$ under the special decomposition $d_nS$ if and only if for some $\alpha \in \{0,1\}$ there exists an $M^{\alpha}$-covering for $S$ under some special decomposition $I(d_nS)$.

Therefore, our goal is to find an $I$-transformation of the decomposition $d_nS$ such that its $\alpha$-domain covers the set $S$. That is, we will search for ordered pairs included in the special decomposition $d_nS$ such that as a result of permutations of their components, the $\alpha$-domain of the resulting decomposition will cover the set $S$.

Let a special decomposition $d_nS$ of the set $S$ be given such that $\{e_{q_1}, \ldots, e_{q_l}\} = S \setminus M^{\alpha}$.

*Definition* 3.1. (i) We say that the subset $M_i^{\alpha}$ is an immediate $M^{\alpha}$-replaceable subset under the special decomposition $d_nS$ if $M^{\alpha} \subseteq (i)M^{\alpha}$.

(ii) We say that the subset $M_i^{\alpha}$ is an $M^{\alpha}$-replaceable subset under the decomposition $d_nS$, if it is an immediate $M^{\alpha}$-replaceable subset or, if $d_nS$ contains ordered pairs

$$(M_{i_1}^{\alpha},\ M_{i_1}^{1-\alpha}), \ldots, (M_{i_k}^{\alpha},\ M_{i_k}^{1-\alpha}),$$

such that

$$M^{\alpha} \subseteq (i, i_1, \ldots, i_k)\ M^{\alpha}.$$

If $d_nS$ does not contain ordered pairs that ensure $M^{\alpha}$-replaceability of the subset $M_i^{\alpha}$, then we say that $M_i^{\alpha}$ is not an $M^{\alpha}$-replaceable subset under the decomposition $d_nS$.

(iii) If the subset $M_i^{\alpha}$ is $M^{\alpha}$-replaceable by ordered pairs

$$(M_i^{\alpha}, M_i^{1-\alpha}), (M_{i_1}^{\alpha}, M_{i_1}^{1-\alpha}), \ldots, (M_{i_k}^{\alpha}, M_{i_k}^{1-\alpha}),$$

then the permutation of components of these ordered pairs will be called an $M^{\alpha}$-replacement of the subset $M_i^{\alpha}$.

(iv) We say that the $M^{\alpha}$-replacement of a subset $M_i^{\alpha}$ is associated with an element $e$, if

$$(e \notin M^{\alpha})\ \&\ (e \in M_i^{1-\alpha})$$

and $M_i^{\alpha}$ is an $M^{\alpha}$-replaceable subset. We also say that the subset $M_i^{\alpha}$ is $M^{\alpha}$-replaceable associated with the element $e$.

(v) We say that an element $e \in S \backslash M^{\alpha}$ is an $M^{\alpha}$-reachable element if there is an ordered pair $(M_i^{\alpha}, M_i^{1-\alpha})$ such that $e \in M_i^{1-\alpha}$ and $M_i^{\alpha}$ is an $M^{\alpha}$-replaceable subset.

(vi) We say that the set $\{e_{q_1}, \ldots, e_{q_r}\} \subseteq S \backslash M^\alpha$ is $M^\alpha$-reachable under the decomposition $d_nS$, if $d_nS$ contains ordered pairs, $(M^\alpha_{i_1}, M^{1-\alpha}_{i_1}), \ldots, (M^\alpha_{i_k}, M^{1-\alpha}_{i_k})$, such that

$$M^\alpha \cup \{e_{q_1}, \ldots, e_{q_r}\} \subseteq (i_1, \ldots, i_k)\, M^\alpha.$$

If $d_nS$ does not contain ordered pairs that ensure $M^\alpha$-reachability the set $\{e_{q_1}, \ldots, e_{q_r}\}$ then we say that the set $\{e_{q_1}, \ldots, e_{q_r}\}$ is not $M^\alpha$-reachable under the decomposition $d_nS$.

The $M^\alpha$-replacement of the subset is actually an $I$-transformation of the given decomposition. Therefore, according to the Lemma 1.6 in [13], for any special decomposition, as a result of any replacement, we obtain a new special decomposition.

*Theorem* 3.2. If $d_nS$ is a special decomposition of the set $S$, and the set

$$c_nS = \{ M_1^{\alpha_1}, \ldots, M_i^{\alpha_i}, \ldots, M_n^{\alpha_n}\}$$

is a special covering for the set $S$ for some $\alpha_1, \ldots, \alpha_n \in \{0,1\}$, then for any $i \in \{1, \ldots, n\}$, the subset $M_i^{1-\alpha_i}$ is an $M^{1-\alpha_i}$-replaceable subset.

*Proof.* Obviously, for any $i \in \{1, \ldots, n\}$, if $M^{1-\alpha_i} \subseteq (i)\, M^{1-\alpha_i}$, then the subset $M_i^{1-\alpha_i}$ is an immediate $M^{1-\alpha_i}$-replaceable subset.

Let for some $i \in \{1, \ldots, n\}$ the subset $M_i^{1-\alpha_i}$ contains elements that are not included in other subsets of the (1-$\alpha_i$)-domain, and therefore $M_i^{1-\alpha_i}$ is not immediate replaceable subset.

Since the set

$$c_nS = \{ M_1^{\alpha_1}, \ldots, M_i^{\alpha_i}, \ldots, M_n^{\alpha_n}\}$$

is a special covering for the set $S$, then according to Proposition 1.4 and Corollary 1.4.1 in [13], for any $e \in M_i^{1-\alpha_i}$ there is a subset $M_j^{\alpha_j}$ such that

$$(M_j^{\alpha_j} \in c_nS)\ \&\ (j \neq i)\ \&\ (e \in M_j^{\alpha_j}).$$

Therefore, there are subsets $M_{i_1}^{\tau_1}, \ldots, M_{i_k}^{\tau_k}$, such that

$$(\{M_{i_1}^{\tau_1}, \ldots, M_{i_k}^{\tau_k}\} \subseteq c_nS)\ \&\ (M_i^{1-\alpha_i} \subseteq \cup_{j=1}^{k} M_{i_j}^{\tau_j}).$$

Since, by the assumption, $M_i^{1-\alpha_i}$ is not immediate replaceable subset and $c_nS$ is a special covering for the set $S$, then some of the elements $M_i^{1-\alpha_i}$ are found in the $\alpha_i$-domain.

It means, for some $\{\beta_1, \ldots, \beta_l\} \subseteq \{\tau_1, \ldots, \tau_k\}$,

$$\{M_{j_1}^{\beta_1}, \ldots, M_{j_l}^{\beta_l}\} \subseteq \{M_{i_1}^{\tau_1}, \ldots, M_{i_k}^{\tau_k}\} \text{ and } \{ M_{j_1}^{\beta_1}, \ldots, M_{j_l}^{\beta_l}\} \subseteq sM^{\alpha_i}.$$

That is, $(\{M_{j_1}^{\beta_1}, \ldots, M_{j_l}^{\beta_l}\} \subseteq c_nS)\ \&\ (\{M_{j_1}^{\beta_1}, \ldots, M_{j_l}^{\beta_l}\} \subseteq sM^{\alpha_i})$.

Obviously, performing permutations of the components of the following ordered pairs,

$$(M_i^{\alpha_i}, M_i^{1-\alpha_i}), (M_{j_1}^{\beta_1}, M_{j_1}^{1-\beta_1}), \ldots, (M_{j_l}^{\beta_l}, M_{j_l}^{1-\beta_l}),$$

the subsets $M_{j_1}^{\beta_1}, \ldots, M_{j_l}^{\beta_l}$ appear in the (1-$\alpha_i$)-domain. If some of the subsets $M_{j_1}^{1-\beta_1}, \ldots, M_{j_l}^{1-\beta_l}$ contain elements that are not included in other subsets in the (1-$\alpha_i$)-domain, then by the same reasoning as for the subset $M_i^{1-\alpha_i}$, we find the corresponding ordered pairs and perform permutations of their components. Thus, the (1-$\alpha_i$)-domain of the special decomposition does not lose elements. So, the subset $M_i^{1-\alpha_i}$ is $M^{1-\alpha_i}$-replaceable. $\nabla$

*Corollary* 3.2.1. Let a special decomposition $d_nS$ of the set $S$ be given such that some element $e$ of the set $S$ is not included in the $\alpha$-domain of this decomposition, $e \notin M^{\alpha}$.

If $e$ is not an $M^{\alpha}$-reachable element, then there is no special covering of the set $S$ under the decomposition of $d_nS$.

*Proof.* If under the given conditions there would be a special covering for the set $S$, then there should be a subset $M_i^{1-\alpha}$ such that

$$(M_i^{1-\alpha} \in c_nS) \ \& \ (e \in M_i^{1-\alpha}).$$

But this means that the conditions of Theorem 3.2 are satisfied for the subset $M_i^{\alpha}$. Thus, the subset $M_i^{\alpha_i}$ will be $M^{\alpha}$-replaceable. And this is contradiction. ∇

*Theorem* 3.3. If there exists a special covering for the set $S$ under the decomposition $d_nS$, then for any $i \in \{1, \ldots, n\}$ and $\alpha \in \{0,1\}$, the following is true:

the subset $M_i^{\alpha}$ is $M^{\alpha}$-replaceable or the subset $M_i^{1-\alpha}$ is $M^{1-\alpha}$-replaceable.

*Proof.* Let for some $\alpha_1, \alpha_2, \ldots, \alpha_n$, the set

$$c_nS = \{ M_1^{\alpha_1}, \ldots, M_i^{\alpha_i}, \ldots, M_n^{\alpha_n}\} \ \ (\alpha_i \in \{0,1\})$$

be a special covering for the set $S$. For each $i \in \{1, \ldots, n\}$ and for some $\alpha \in \{0,1\}$ we have

$$\alpha_i = \alpha \ \text{ or } \ \alpha_i = 1\text{-}\alpha.$$

Then, since the conditions of theorem 3.2 are satisfied, we have:

for $\alpha_i = \alpha$, there is an $M^{1-\alpha}$-replacement of the subset $M_i^{1-\alpha}$,

for $\alpha_i = 1\text{-}\alpha$ there is an $M^{\alpha}$-replacement of the subset $M_i^{\alpha}$. ∇

An important problem arises in studying the existence of a special covering for a set.

Let $\{e_{q_1}, \ldots, e_{q_l}\} = S \setminus M^{\alpha}$. The elements of the set $\{e_{q_1}, \ldots, e_{q_l}\}$ can occur in different subsets in the (1-α)-domain of the decomposition $d_nS$. So, to find a special covering for the set $S$, we can do one of the following procedures:

1) search for ordered pairs of the decomposition $d_nS$, which will ensure the $M^{\alpha}$-reachability of the set $\{e_{q_1}, \ldots, e_{q_l}\}$.

2) sequential search for $M^{\alpha}$-reachability of each element of the set $\{e_{q_1}, \ldots, e_{q_l}\}$, that is:

- search for ordered pairs that will ensure the $M^{\alpha}$-reachability of the element $e_{q_1}$ under the decomposition $d_nS$, and, having found them, permute their components. The result will be a new decomposition.

- search for ordered pairs, that will ensure the $M^{\alpha}$-reachability of the element $e_{q_2}$ under the new decomposition, and, having found them, permute their components, and so on.

Therefore, it is important to clarify the following:

- whether the existence of a special covering of the set $S$ depends on how we study the $M^{\alpha}$-reachability of the elements of the set $\{e_{q_1}, \ldots, e_{q_l}\}$.

- in the sequential search for $M^{\alpha}$-reachability of elements of the set $\{e_{q_1}, \ldots, e_{q_l}\}$, whether the existence of a special covering depends on the order in which these elements are considered.

*Definition* 3.4. Let $d_nS$ be a special decomposition of the set $S$.

We define the decomposition $d_nS^{\alpha}(e_{i_1}, \ldots, e_{i_l})$ for an ordered set $\{e_{i_1}, \ldots, e_{i_l}\} \subseteq S$ and fixed $\alpha \in \{0,1\}$ as follows:

(i) if $e_{i_1} \in M^{\alpha}$, then $d_nS^{\alpha}(e_{i_1})$ is identical with the decomposition $d_nS$.

(ii) let $e_{i_1} \notin M^{\alpha}$. We consider the following cases:

- if there is an ordered pair, $(M_i^{\alpha}, M_i^{1-\alpha}) \in d_nS$, such that $e_{i_1} \in M_i^{1-\alpha}$ and $M_i^{\alpha}$ is an $M^{\alpha}$-replaceable subset, then $d_nS^{\alpha}(e_{i_1})$ is the decomposition obtained as a result of $M^{\alpha}$-replacement of the subset $M_i^{\alpha}$ under the decomposition $d_nS$.

- if $e_{i_1}$ is not an $M^{\alpha}$-reachable element under the decomposition $d_nS$, we assume that $d_nS^{\alpha}(e_{i_1}) = \emptyset$.

(iii) let $d_nS^{\alpha}(e_{i_1}, \ldots, e_{i_k}) \neq \emptyset$ for some $1 \leq k < l$.

We consider the following cases:

1) if $e_{i_{k+1}}$ is included in the $\alpha$-domain of the decomposition $d_nS^{\alpha}(e_{i_1}, \ldots, e_{i_k})$, then $d_nS^{\alpha}(e_{i_1}, \ldots, e_{i_k}, e_{i_{k+1}})$ is identical with the decomposition $d_nS^{\alpha}(e_{i_1}, \ldots, e_{i_k})$.

2) let $e_{i_{k+1}}$ is not included in the $\alpha$-domain of the decomposition $d_nS^{\alpha}(e_{i_1}, \ldots, e_{i_k})$.

- if there is an ordered pair, $(M_i^{\alpha}, M_i^{1-\alpha}) \in d_nS^{\alpha}(e_{i_1}, \ldots, e_{i_k})$, such that $e_{i_{k+1}} \in M_i^{1-\alpha}$ and $M_i^{\alpha}$ is an $M^{\alpha}$-replaceable subset under the decomposition $d_nS^{\alpha}(e_{i_1}, \ldots, e_{i_k})$, then we assume that $d_nS^{\alpha}(e_{i_1}, \ldots, e_{i_k}, e_{i_{k+1}})$ is the decomposition that is obtained as a result of $M^{\alpha}$-replacement procedure of $M_i^{\alpha}$ under the decomposition $d_nS^{\alpha}(e_{i_1}, \ldots, e_{i_k})$.

- if $e_{i_{k+1}}$ is not an $M^{\alpha}$-reachable element under the decomposition $d_nS^{\alpha}(e_{i_1}, \ldots, e_{i_k})$, we assume that $d_nS^{\alpha}(e_{i_1}, \ldots, e_{i_k}, e_{i_{k+1}}) = \emptyset$.

(iv) if for some $1 \leq k < l$, $d_nS^{\alpha}(e_{i_1}, \ldots, e_{i_k}) = \emptyset$, we assume that

$$d_nS^{\alpha}(e_{i_1}, \ldots, e_{i_k}, e_{i_{k+1}}) = \emptyset.$$

*Theorem* 3.5. Let the special decomposition $d_nS$ of the set $S$ is given such that

$$\{e_{q_1}, \ldots, e_{q_l}\} = S \setminus M^{\alpha}.$$

(i) if there is a special covering for the set $S$ under the decomposition $d_nS$, then for any ordered set $\{e_{j_1}, \ldots, e_{j_l}\}$ obtained as a result of an arbitrary permutation of elements of the ordered set $\{e_{q_1}, \ldots, e_{q_l}\}$, the following is true:

- $d_nS^{\alpha}(e_{j_1}, \ldots, e_{j_l}) \neq \emptyset$,

- the $\alpha$-domain of the decomposition $d_nS^{\alpha}(e_{j_1}, \ldots, e_{j_l})$ is an $M^{\alpha}$-covering for the set $S$.

(ii) if $d_nS^{\alpha}(e_{j_1}, \ldots, e_{j_k}) = \emptyset$ for some $\{e_{j_1}, \ldots, e_{j_k}\} \subseteq \{e_{q_1}, \ldots, e_{q_l}\}$, then there does not exist a special covering for the set $S$ under the decomposition $d_nS$.

*Proof.* (i) Suppose, for some $\alpha_1, \ldots, \alpha_n$, the set

$$c_nS = \{M_1^{\alpha_1}, \ldots, M_i^{\alpha_i}, \ldots, M_n^{\alpha_n}\}$$

is a special covering for the set $S$ under the decomposition $d_nS$.

$$S \setminus M^{\alpha} = \{e_{q_1}, \ldots, e_{q_l}\},$$

implies that $c_nS \neq sM^{\alpha}$ and $\{e_{q_1}, \ldots, e_{q_l}\} \subseteq M^{1-\alpha}$.

Since $c_nS$ is a special covering for the set $S$ and $\{e_{q_1}, \ldots, e_{q_l}\} = S \setminus M^\alpha$, then the elements $e_{q_1}, \ldots, e_{q_l}$ belong to some subsets of the (1-$\alpha$)-domain, and each of them also belongs to some subset included in $c_nS$. Therefore, for any $e_j \notin M^\alpha$ , there exists a subset included in $c_nS$, which contains the element $e_j$. Suppose the elements of the set $\{e_{q_1}, \ldots, e_{q_l}\}$ are included in the subsets

$$M_{i_1}^{1-\alpha}, \ldots, M_{i_k}^{1-\alpha} \in c_nS.$$

It is easy to see that for any $M_{i_j}^{\alpha} \in \{M_{i_1}^{1-\alpha}, \ldots, M_{i_k}^{1-\alpha}\}$ the conditions of Theorem 3.2 are satisfied. So, the subset $M_{i_j}^{\alpha}$ is $M^\alpha$-replaceable. Let $e_{j_1} \in M_{i_1}^{1-\alpha}$. Performing the $M^\alpha$-replacement of the subset $M_{i_1}^{\alpha}$, we obtain the special decomposition $d_nS^\alpha(e_{j_1}) \neq \emptyset$.

Since the transition from the decomposition $d_nS$ to the decomposition $d_nS^\alpha(e_{j_1})$ is an $I$-transformation, then according to Lemma 1.6 in [13], there is a special covering for $S$ under the decomposition $d_nS^\alpha(e_{j_1})$.

If as a result of $M^\alpha$-replacement associated with the element $e_{j_1}$, the element $e_{j_2}$ also moved to the $\alpha$-domain, then

$$d_nS^\alpha(e_{j_1}) = d_nS^\alpha(e_{j_1}, e_{j_2}) \text{ and } d_nS^\alpha(e_{j_1}, e_{j_2}) \neq \emptyset.$$

If still $e_{j_2} \notin M^\alpha$, then by the same reasoning, as in the case of the element $e_{j_1}$, according to the propositions 1.4 and 1.4.1 in [13], there exists a subset $M_j^{1-\alpha}$ such, that

$$(M_j^{1-\alpha} \in c_nS) \ \& \ (e_{j_2} \in M_j^{1-\alpha}).$$

And, according to the Theorem 3.2, the subset $M_j^\alpha$ is $M^\alpha$-replaceable. Since the replacement is associated with the element $e_{j_2}$, then performing the $M^\alpha$-replacement of the subset $M_j^\alpha$, we obtain the following:

- $d_nS^\alpha(e_{j_1}, e_{j_2}) \neq \emptyset$,
- $d_nS^\alpha(e_{j_1}, e_{j_2})$ is a special decomposition,
- there is a special covering for the set $S$ under the decomposition $d_nS^\alpha(e_{j_1}, e_{j_2})$.

Repeating the similar procedure, no more than $l$ times, we obtain:

- $d_nS^\alpha(e_{j_1}, \ldots, e_{j_l}) \neq \emptyset$,
- $d_nS^\alpha(e_{j_1}, \ldots, e_{j_l})$ is a special decomposition,
- there is a special covering for the set $S$ under the decomposition $d_nS^\alpha(e_{j_1}, \ldots, e_{j_l})$.

Since as a result of formation of decomposition $d_nS^\alpha(e_{j_1}, \ldots, e_{j_l})$ all elements of the set $\{e_{j_1}, \ldots, e_{j_l}\}$ move to the $\alpha$-domain, then, obviously, the $\alpha$-domain of the decomposition $d_nS^\alpha(e_{j_1}, \ldots, e_{j_l})$ is an $M^\alpha$-covering for the set $S$.

(ii) Suppose now, that $d_nS^\alpha(e_{j_1}, \ldots, e_{j_k}) = \emptyset$ for some set $\{e_{j_1}, \ldots, e_{j_k}\} \subseteq \{e_{q_1}, \ldots, e_{q_l}\}$, and show that there does not exist a special covering for the set $S$ under decomposition $d_nS$.

Recall, that $d_nS^\alpha(e_{j_1}, \ldots, e_{j_k}) = \emptyset$ in the following cases:

- $d_nS^\alpha(e_{j_1}) = \emptyset$.

This means, there is no $M^\alpha$-replaceable subset associated with the element $e_{j_1}$, under the decomposition $d_nS$. Therefore, there does not exist a special covering for the set $S$.

- for some element $e_{j_r}$ ( $1 \leq r < k$ ),

$$d_nS^\alpha(e_{j_1}, \ldots, e_{j_r}) \neq \emptyset \ \& \ d_nS^\alpha(e_{j_1}, \ldots, e_{j_r}, e_{j_{r+1}}) = \emptyset.$$

This means that $d_nS^{\alpha}(e_{j_1}, \ldots, e_{j_r})$ is a special decomposition of the set $S$ and there is no $M^{\alpha}$-replaceable subset associated with the element $e_{j_{r+1}}$ under the decomposition

$$d_nS^{\alpha}(e_{j_1}, \ldots, e_{j_r}).$$

According to the Theorem 3.2.1 there cannot exist a special covering for the set $S$ under the special decomposition $d_nS^{\alpha}(e_{j_1}, \ldots, e_{j_r})$.

Since the decomposition $d_nS^{\alpha}(e_{j_1}, \ldots, e_{j_r})$ is obtained from the decomposition $d_nS$ as a result of an $I$-transformation, according to Lemma 1.6 in [13], the decomposition $d_nS$ does not allow a special covering for the set $S$. ∇

*Corollary* 3.5.1. If a special decomposition $d_nS$ of the set $S$ is given such that

$$\{e_{q_1}, \ldots, e_{q_l}\} = S \setminus M^{\alpha},$$

then the set $\{e_{q_1}, \ldots, e_{q_l}\}$ is $M^{\alpha}$-reachable under the decomposition $d_nS$ if and only if

$$d_nS^{\alpha}(e_{q_1}, \ldots, e_{q_l}) \neq \emptyset.$$

*Proof.* Let $\{e_{q_1}, \ldots, e_{q_l}\}$ be an $M^{\alpha}$-reachable set under the decomposition $d_nS$. This means that $d_nS$ contains ordered pairs

$$(M_{i_1}^{\alpha}, M_{i_1}^{1-\alpha}), \ldots, (M_{i_k}^{\alpha}, M_{i_k}^{1-\alpha})$$

such that as a result of permutations of the components of these pairs, we will have

$$M^{\alpha} \cup \{e_{q_1}, \ldots, e_{q_r}\} \subseteq (i_1, \ldots, i_k)\, M^{\alpha}.$$

That is, $\alpha$-domain of the decomposition $(i_1, \ldots, i_k)I(d_nS)$ includes all elements of the $\alpha$-domain of the decomposition $d_nS$ and the elements of the set $\{e_{q_1}, \ldots, e_{q_r}\}$. Therefore, $(i_1, \ldots, i_k)\, M^{\alpha}$ is an $M^{\alpha}$-covering for the set $S$ under the decomposition $(i_1, \ldots, i_k)I(d_nS)$. Thus, according to Lemma 1.6 in [13], there is a special covering for $S$ under the decomposition $d_nS$, and according to Theorem 3.5 $d_nS^{\alpha}(e_{q_1}, \ldots, e_{q_l}) \neq \emptyset$.

At the other hand, if $d_nS^{\alpha}(e_{q_1}, \ldots, e_{q_l}) \neq \emptyset$, then, obviously, it can be obtained from the decomposition $d_nS$ by permuting the components of some ordered pair involved in the procedure for forming the decomposition $d_nS^{\alpha}(e_{q_1}, \ldots, e_{q_l})$.

This means that these ordered pairs provide $M^{\alpha}$-reachability of the set $\{e_{q_1}, \ldots, e_{q_l}\}$ under the decomposition $d_nS$. ∇

The result of Theorem 3.5 is important in that having a set of elements $\{e_{q_1}, \ldots, e_{q_l}\}$ that are not included in the $\alpha$-domain of the given decomposition, we can search for the $M^{\alpha}$-reachability of each of these elements, considering them in any order.

At the same time, important questions remain open in Definition 3.4:

How to find ordered pairs that ensure the reachability of a given element, or how conclude that such ordered pairs do not exist.

Therefore, our further research will be aimed at finding a way and determining the actions that will lead to the assertion or denial of the possibility of replaceability of the subset under consideration for a given decomposition.

## 4. Pointing Graphs

Let $d_nS$ be a special decomposition of the set $S = \{e_1, e_2, \ldots, e_m\}$ such that

$$\{e_{q_1}, \ldots, e_{q_l}\} = S \setminus M^{\alpha}.$$

To find a special covering for the set $S$ under the special decomposition $d_nS$, we need procedures for searching for ordered pairs included in the set $d_nS$, that ensure $M^{\alpha}$-reachability of the elements of the set $\{e_{q_1}, \ldots, e_{q_l}\}$.

Therefore, we introduce the concept of a graph generated by a given special decomposition.

Using such graphs, it will be convenient to describe the procedures for finding the required ordered pairs in a given decomposition, if they exist. We will call such a graph an $M^{\alpha}$-pointing graph.

*Definition* 4.1. The subset $M_i^{\alpha}$ will be called an $M^{\alpha}$-single subset with respect to some elements $e_{j_1}, \ldots, e_{j_l}$ under the special decomposition $d_nS$, if

- $\{e_{j_1}, \ldots, e_{j_l}\} \subseteq M_i^{\alpha}$,
- none of the elements $e_{j_1}, \ldots, e_{j_l}$ belongs to other subsets of the $\alpha$-domain.

If no uncertainty occurs, we will sometimes skip the enumeration of elements, with respect to which the subset is $M^{\alpha}$-single.

*Definition* 4.2. (i) The procedure for permuting the components of any ordered pair will be called a replacement step or, briefly, a step. It will also be denoted as the corresponding ordered pair, $(M_i^{\alpha}, M_i^{1-\alpha})$.

(ii) we will say that the replacement step $(M_i^{\alpha}, M_i^{1-\alpha})$ leads to the replacement step $(M_j^{\alpha}, M_j^{1-\alpha})$ to reach the element $e \in S$, if:

- $(e \in M_i^{\alpha})$ and $M_i^{\alpha}$ is an $M^{\alpha}$-single subset with respect to the element $e$,
- $(e \in M_j^{1-\alpha})$.

(iii) the replacement step $(M_j^{\alpha}, M_j^{1-\alpha})$ is called an obligatory step associated with an element $e \in S$, if:

- $(e \notin M^{\alpha})$ & $(e \in M_j^{1-\alpha})$,
- $M_j^{1-\alpha}$ is an $M^{1-\alpha}$-single subset with respect to the element $e$.

(iv) the replacement step $(M_j^{\alpha}, M_j^{1-\alpha})$ is called a possible step associated with an element $e \in S$, if:

- $(e \notin M^{\alpha})$ & $(e \in M_j^{1-\alpha})$,
- there are also other subsets $M_{j_1}^{1-\alpha}, \ldots, M_{j_k}^{1-\alpha}$ in the (1-$\alpha$)-domain such that

$$(e \in M_{j_1}^{1-\alpha}) \ \& \ldots \& \ (e \in M_{j_k}^{1-\alpha}).$$

(v) the replacement step $(M_i^{\alpha}, M_i^{1-\alpha})$ is called a useless step associated with an element $e \in S$, if $M_i^{\alpha}$ is an $M^{\alpha}$-single subset with respect to the element $e$ and $e \notin M^{1-\alpha}$.

(vi) the replacement step $(M_j^{\alpha}, M_j^{1-\alpha})$ is called a final step, if the subset $M_j^{\alpha}$ is an immediate $M^{\alpha}$-replaceable subset.

*Remark* 4.3. If $M_i^{\alpha}$ is the only subset in the $\alpha$-domain containing the element $e$, then after the replacement step $(M_i^{\alpha}, M_i^{1-\alpha})$, the $\alpha$-domain loses $e$ and, to return it back, it is necessary to perform the step $(M_j^{\alpha}, M_j^{1-\alpha})$.

*Definition* 4.4. (i) Let $d_nS$ be a special decomposition such that $\{e_{q_1}, \ldots, e_{q_l}\} \subseteq S \setminus M^{\alpha}$.

The subsets $M_{j_1}^{1-\alpha}, \ldots, M_{j_p}^{1-\alpha}$ of the (1-$\alpha$)-domain of the decomposition $d_nS$ will be called main subsets associated with the set $\{e_{q_1}, \ldots, e_{q_l}\}$, if:

- $M_{j_k}^{1-\alpha} \in \{M_{j_1}^{1-\alpha}, \ldots, M_{j_p}^{1-\alpha}\}$ implies $M_{j_k}^{1-\alpha} \cap \{e_{q_1}, \ldots, e_{q_l}\} \neq \emptyset$,
- none of the elements $e_{q_1}, \ldots, e_{q_l}$ is included in other subsets of the (1-$\alpha$)-domain.

(ii) If $M_{j_1}^{1-\alpha}, \ldots, M_{j_p}^{1-\alpha}$ are the main subsets associated with the set $\{e_{q_1}, \ldots, e_{q_l}\}$, then the set

$$R_M = \{(M_{j_1}^{\alpha}, M_{j_1}^{1-\alpha}), \ldots, (M_{j_p}^{\alpha}, M_{j_p}^{1-\alpha})\}$$

will be called a set of main replacement steps associated with the set $\{e_{q_1}, \ldots, e_{q_l}\}$.

4.4.1. The set of ordered pairs,

$$R = \{(M_{i_1}^{\alpha}, M_{i_1}^{1-\alpha}), \ldots, (M_{i_l}^{\alpha}, M_{i_l}^{1-\alpha})\} \subseteq d_nS,$$

will be called the description of an $M^{\alpha}$-replaceability procedure associated with the set $\{e_{q_1}, \ldots, e_{q_l}\}$, or, briefly, a replaceability procedure, if:

- $R_M \subseteq R$,
- for any replacement step $(M_k^{\alpha}, M_k^{1-\alpha}) \in R \setminus R_M$, there is another replacement step $(M_j^{\alpha}, M_j^{1-\alpha}) \in R$ such that $(M_j^{\alpha}, M_j^{1-\alpha})$ leads to $(M_k^{\alpha}, M_k^{1-\alpha})$ to reach some element,
- if some replacement step, $(M_j^{\alpha}, M_j^{1-\alpha}) \in R$, leads to another step $(M_k^{\alpha}, M_k^{1-\alpha})$, then $(M_k^{\alpha}, M_k^{1-\alpha}) \in R$.

4.4.2. For some main replacement step $(M_i^{\alpha}, M_i^{1-\alpha}) \in R_M$, the set

$$R = \{(M_i^{\alpha}, M_i^{1-\alpha}), (M_{i_1}^{\alpha}, M_{i_1}^{1-\alpha}), \ldots, (M_{i_l}^{\alpha}, M_{i_l}^{1-\alpha})\}$$

is called the description of an $M^{\alpha}$-replaceability procedure started form the step $(M_i^{\alpha}, M_i^{1-\alpha})$, if

- for any replacement step

$$(M_k^{\alpha}, M_k^{1-\alpha}) \in R \setminus \{(M_i^{\alpha}, M_i^{1-\alpha})\},$$

there is another replacement step $(M_j^{\alpha}, M_j^{1-\alpha}) \in R$ such that $(M_j^{\alpha}, M_j^{1-\alpha})$ leads to $(M_k^{\alpha}, M_k^{1-\alpha})$,

- if some replacement step $(M_j^{\alpha}, M_j^{1-\alpha}) \in R$ leads to another step $(M_k^{\alpha}, M_k^{1-\alpha})$, then $(M_k^{\alpha}, M_k^{1-\alpha}) \in R$.

4.4.3. If the set $R$ is the description of an $M^{\alpha}$-replaceability procedure, then the permutation of the components of ordered pairs included in the set $R$ will be called the replacement procedure described by the set $R$.

*Proposition* 4.5. If $\{e_{q_1}, \ldots, e_{q_l}\}$ is an $M^{\alpha}$-reachable set under the special decomposition $d_nS$, and the set of ordered pairs,

$$R = \{(M^{\alpha}_{r_1},\ M^{1-\alpha}_{r_1}), \ldots, (M^{\alpha}_{r_l},\ M^{1-\alpha}_{r_l})\},$$

is the description of an $M^{\alpha}$-replaceability procedure associated with the set $\{e_{q_1}, \ldots, e_{q_l}\}$, then the ordered pairs necessary to ensure an $M^{\alpha}$-reachability of the set $\{e_{q_1}, \ldots, e_{q_l}\}$ are included in the set $R$.

*Proof.* Assume that

$$R_1 = \{(M^{\alpha}_{i_1}, M^{1-\alpha}_{i_1}), \ldots, (M^{\alpha}_{i_k}, M^{1-\alpha}_{i_k})\}$$

is the set of ordered pairs needed to provide an $M^{\alpha}$-reachability of the set $\{e_{q_1}, \ldots, e_{q_l}\}$.

By definition, $R$ contains the set of main replacement steps associated with the set of elements $\{e_{q_1}, \ldots, e_{q_l}\}$, i.e.

$$R_M = \{(M^{\alpha}_{j_1}, M^{1-\alpha}_{j_1}), \ldots, (M^{\alpha}_{j_p}, M^{1-\alpha}_{j_p})\} \subseteq R.$$

Obviously, also $R_M \subseteq R_1$.

If the set $R$ consists only of the main replacement steps, then all of them are immediate $M^{\alpha}$-replaceable subsets and, therefore, are the final replacement steps.

Obviously, $R_1$ also consists only of the main replaceability steps. Therefore, $R_1 \subseteq R$.

Let the set $R_M$ contains an ordered pair, $(M^{\alpha}_{j}, M^{1-\alpha}_{j}) \in R_M$, such that $M^{\alpha}_{j}$ is an $M^{\alpha}$-single subset with respect to some elements.

Since $R_M \subseteq R_1$ and the set $R_1$ provides an $M^{\alpha}$-reachability of the set $\{e_{q_1}, \ldots, e_{q_l}\}$, then after the permutation of components of the ordered pair $(M^{\alpha}_{j},\ M^{1-\alpha}_{j})$, the $\alpha$-domain loses elements.

This means that $R_1$ contains ordered pairs, after the permutations of the components of which the lost elements are returned back to $\alpha$-domain. At the same time, these ordered pairs are also included in the set $R$, because of following:

- the replaceability procedure $R$ begins with the main replacement steps,
- if some replacement step $(M^{\alpha}_{j}, M^{1-\alpha}_{j}) \in R$ leads to another step $(M^{\alpha}_{r}, M^{1-\alpha}_{r})$, then

$$(M^{\alpha}_{r}, M^{1-\alpha}_{r}) \in R.$$

Using this reasoning, we conclude that if the $\alpha$-domain loses elements after any replacement step, then the set $R$ contains all the necessary replacement steps to return these elements.

Thus, if an ordered pair is included in the set $R_1$, it is also included in the set $R$.

Therefore $R_1 \subseteq R$. ∇

If, under the decomposition $d_nS$, none of the sets $sM^{\alpha}$ and $sM^{1-\alpha}$ covers the set $S$, we proceed to investigate $M^{\alpha}$-replaceability procedure associated with the elements not included in the $\alpha$-domain.

Our goal is to find out if these elements are $M^{\alpha}$-reachable.

To do this, it is convenient to represent the description of the replaceability procedure as a directed graph with labeled edges.

To construct a graph, the set of ordered pairs included in the set $R$ will be considered as the set of vertices of the graph, which will be denoted by $V$. We will use the notation $v_i$ for the vertex corresponding to the ordered pair $(M_i^{\alpha}, M_i^{1-\alpha})$.

If a replacement step $(M_i^{\alpha}, M_i^{1-\alpha})$ leads to a replacement step $(M_j^{\alpha}, M_j^{1-\alpha})$, we consider the ordered pair of vertises $(v_i, v_j)$ as a directed edge of the graph. We denote by $E$ the set of edges of the graph.

The labels of the edges are defined as follows:

If the replacement step $(M_i^{\alpha}, M_i^{1-\alpha})$ leads to the replacement step $(M_j^{\alpha}, M_j^{1-\alpha})$ to reach an element $e \in S$, and $(M_j^{\alpha}, M_j^{1-\alpha})$ is an obligatory step associated with the element $e$, then the edge $(v_i, v_j)$ will be called a conjunctive edge associated with the element $e$. We add the label "&$e$" to this edge and denote it as follows:

$$( v_i, \&e, v_j).$$

If the replacement step $(M_i^{\alpha}, M_i^{1-\alpha})$ leads to the replacement step $(M_j^{\alpha}, M_j^{1-\alpha})$ to reach an element $e \in S$, and $(M_j^{\alpha}, M_j^{1-\alpha})$ is a possible step associated with the element $e$, then the edge $(v_i, v_j)$ will be called a disjunctive edge associated with the element $e$. We add the label "$\vee e$" to this edge and denote it as follows:

$$(v_i, \vee e, v_j).$$

If the replacement step $(M_j^{\alpha}, M_j^{1-\alpha})$ is a useless step associated with some element $e$, then the corresponding vertex $v_j$ will be called a useless vertex associated with the element $e$.

If the replacement step $(M_j^{\alpha}, M_j^{1-\alpha})$ is a final step, then the corresponding vertex $v_j$ will be called a final vertex.

We will say that the vertex $v_j$ is associated with some element $e \in S$, if the corresponding step $(M_j^{\alpha}, M_j^{1-\alpha})$ is associated with the element $e$.

The vertices $v_{j_1}, \ldots, v_{j_p}$ of the graph under construction will be called main vertices, if

- $\{e_{q_1}, \ldots, e_{q_l}\} \subseteq S \setminus M^{\alpha}$,
- the vertices $v_{j_1}, \ldots, v_{j_p}$ correspond to the ordered pairs

$$(M_{j_1}^{\alpha}, M_{j_1}^{1-\alpha}), \ldots, (M_{j_p}^{\alpha}, M_{j_p}^{1-\alpha}),$$

- the set of ordered pairs

$$\{(M_{j_1}^{\alpha}, M_{j_1}^{1-\alpha}), \ldots, (M_{j_p}^{\alpha}, M_{j_p}^{1-\alpha})\}$$

is the set of the main replacement steps associated with the set $\{e_{q_1}, \ldots, e_{q_l}\}$.

The general scheme of the graph construction procedure consists of the following:

- we determine the main vertices of the graph and form the initial set of graph vertices, that is, $V = \{v_{j_1}, \ldots, v_{j_p}\}$,
- we sequentially examine each vertex $v \in V$ in order to find outgoing edges and new vertices of the graph, if they exist.
- we add the appeared new vertices to the set $V$ and continue studying the other vertices.
- the procedure ends if we have considered all the vertices of the set $V$ and new edges and vertices no longer appear.

4.6. *Graph construction procedure.*

Let us describe the procedure for constructing a pointing graph based on the special decomposition $d_n S$, assuming that:

- $\{e_{q_1}, \ldots, e_{q_l}\} \subseteq S \setminus M^{\alpha}$ and $\{e_{q_1}, \ldots, e_{q_l}\} \neq \emptyset$,
- the subsets $M_{j_1}^{1-\alpha}, \ldots, M_{j_p}^{1-\alpha}$ are main subsets associated with the set $\{e_{q_1}, \ldots, e_{q_l}\}$.

This means that $v_{j_1}, \ldots, v_{j_p}$ will be the main vertices of the graph under construction, therefore, the initial set of vertices of the graph consists of them, $V = \{v_{j_1}, \ldots, v_{j_p}\}$.

Sequentially considering the vertices included in the set $V$, for each vertex $v_i \in V$ corresponding to the replacement step $(M_i^{\alpha}, M_i^{1-\alpha})$, we do the following:

(i) if $M_i^{\alpha}$ is an immediate $M^{\alpha}$-replaceable subset, then there are no edges outgoing from the vertex $v_i$, which means that $v_i$ is a final vertex. So, we mark it as a studied vertex and proceed to consider another vertex of the set $V$.

(ii) if $M_i^{\alpha}$ is an $M^{\alpha}$-single subset with respect to the elements $e_{i_1}, \ldots, e_{i_k}$, then for each $e_{i_j} \in \{e_{i_1}, \ldots, e_{i_k}\}$ we find all subsets in the (1- $\alpha$)-domain containing the element $e_{i_j}$. Let them be the subsets $M_{r_1}^{1-\alpha}, \ldots, M_{r_l}^{1-\alpha}$.

This means that the replacement step $(M_i^{\alpha}, M_i^{1-\alpha})$ leads to the replacement steps

$$(M_{r_1}^{\alpha}, M_{r_1}^{1-\alpha}), \ldots, (M_{r_{l_j}}^{\alpha}, M_{r_l}^{1-\alpha})$$

to reach the element $e_{i_j}$.

Thus, with each element $e_{i_j} \in \{e_{i_1}, \ldots, e_{i_k}\}$, the vertices $v_{r_1}, \ldots, v_{r_l}$ appear, so with the vertex $v_i$ we compose new directed edges of the graph, $(v_i, v_{r_1}), \ldots, (v_i, v_{r_l})$.

We add the vertices $v_{r_1}, \ldots, v_{r_l}$ to the set $V$ if they are not already there. To determine the labels for edges, we do the following:

Let for each $e_{i_j} \in \{e_{i_1}, \ldots, e_{i_k}\}$, $l_j$ be the number of all subsets in the (1-$\alpha$)-domain that contain the element $e_{i_j}$.

(ii.1) if $l_j$ =1, then

- $e_{i_j} \in M_{r_1}^{1-\alpha}$,
- $M_{r_1}^{1-\alpha}$ is $M^{1-\alpha}$-single subset with respect to $e_{i_j}$,
- $(M_{r_1}^{\alpha}, M_{r_1}^{1-\alpha})$ is the only replacement step associated with the element $e_{i_j}$.

So, $(M_{r_1}^{\alpha}, M_{r_1}^{1-\alpha})$ is an obligatory step associated with the element $e_{i_j}$, and therefore, the edge $(v_i, v_{r_1})$ will be a conjunctive edge. We mark this edge by the label $\&e_{i_j}$ and use the notation $(v_i, \&e_{i_j}, v_{r_1})$ for it. We also add this edge to the set $E$.

(ii.2) if $l_j > 1$, then the ordered pairs

$$(M_{r_1}^{\alpha}, M_{r_1}^{1-\alpha}), \ldots, (M_{r_{l_j}}^{\alpha}, M_{r_{l_j}}^{1-\alpha})$$

are possible steps, associated with the element $e_{i_j}$. So, we form the labeled edges

$$(v_i, \vee e_{i_j}, v_{r_1}), \ldots, (v_i, \vee e_{i_j}, v_{r_{l_j}})$$

associated with the element $e_{i_j}$. We add all the formed edges to the set $E$.

(ii.3) if $l_j = 0$ for some $e_{i_j} \in \{e_{i_1}, \ldots, e_{i_k}\}$, then $e_{i_j} \notin M^{1-\alpha}$ and $(M_i^{\alpha}, M_i^{1-\alpha})$ is a useless replacement step associated with the element $e_{i_j}$, which do not lead to another replacement step associated with $e_{i_j}$. So, we mark $v_i$ as a useless vertex.

(ii.4) if for each element $e_{i_j} \in \{e_{i_1}, \ldots, e_{i_k}\}$, all edges associated with $e_{i_j}$ are found, we mark the vertex $v_i$ as the already studied vertex.

(iii) the graph construction procedure will be completed if all the vertices belonging to $V$ have been studied, and new vertices and edges no longer appear.

We will use the notation $G(e_{q_1}, \ldots, e_{q_l})$ for the graph constructed according to this description and call it the graph corresponding to the replaceability procedure associated with set $\{e_{q_1}, \ldots, e_{q_l}\}$ under the special decomposition $d_nS$.

Sometimes, if it does not lead to ambiguity, we will skip listing the elements and use the notation $G$ for the graph, and also call it an $M^{\alpha}$-pointing graph or, simply, a graph.

*Remark* 4.6.1. Suppose that $\{v_{i_1}, \ldots, v_{i_r}\}$ is the set of vertices of the $M^{\alpha}$-pointing graph $G(e_{q_1}, \ldots, e_{q_l})$, which corresponds to the description of the replaceability procedure associated with set of elements $\{e_{q_1}, \ldots, e_{q_l}\}$ under the special decomposition $d_nS$.

Permuting the components of the ordered pairs corresponding to the vertices of the graph, according to Lemma 1.6 in [13], we obtain a new special decomposition $(i_1, \ldots, i_r)I(d_nS)$ of the set $S$.

*The decomposition* $(i_1, \ldots, i_r)I(d_nS)$ *will be called the decomposition corresponding to the vertices of the graph* $G(e_{q_1}, \ldots, e_{q_l})$.

*We will use the notation* $G[v_i]$ *for the graph corresponding to the replaceability procedure started from the main replacement step* $(M_i^{\alpha}, M_i^{1-\alpha})$.

<u>*Proposition*</u> 4.7. If $v_{j_1}, \ldots, v_{j_p}$ are the main vertices of the graph $G(e_{q_1}, \ldots, e_{q_l})$, then for any $v_i \in \{v_{j_1}, \ldots, v_{j_p}\}$ the following is true:

- the graph $G[v_i]$ is a connected graph,
- there is a path from the main vertex to any other vertex of the graph $G[v_i]$.

<u>*Proof*</u>. The only main vertex of the graph $G[v_i]$ is $v_i$, which corresponds to the ordered pair $(M_i^{\alpha}, M_i^{1-\alpha})$, and the construction of the graph is started from it.

Obviously, the proposition is true if $M_i^{\alpha}$ is an immediate $M^{\alpha}$-replaceable subset. Let $M_i^{\alpha}$ it be an $M^{\alpha}$-single subset. Since the vertices and edges of the graph appear when some replacement step corresponding to an already formed vertex leads to another replacement step, we get a connected graph in which there is a path from $v_i$ to any other vertex. ∇

<u>*Proposition*</u> 4.8. Let $G(e_{q_1}, \ldots, e_{q_l})$ be an $M^{\alpha}$-pointing graph corresponding to some $M^{\alpha}$-replaceability procedure under some special decomposition $d_nS$.

Then, for any path in the graph, started from some main vertex, only one of the following statements is true:

- the path ends reaching some final vertex of the graph,
- the path is interrupted, meeting a particular useless vertex of the graph,
- the path turns into a cycle containing at most $n$-1 successive edges.

<u>*Proof*</u>. The path in the graph is formed as a result of successive replacement steps, each of which leads to the next step. Therefore, having reached the vertex corresponding to the final replacement step, the path ends.

Let $(M_i^{\alpha}, M_i^{1-\alpha})$ be a useless replacement step associated with some element $e \in S$. If the subset $M_i^{\alpha}$ does not contain other elements with respect to which it is $M^{\alpha}$-single, then the path in the graph, reaching the vertex corresponding to $(M_i^{\alpha}, M_i^{1-\alpha})$, is interrupted.

Any pointing graph contains at most $n$ vertices. This means that if neither the final vertex nor the corresponding useless vertex occurs along the path when passing ($n$ - 1) successive edges, then the next edge outgoing from the last vertex will lead to one of its ancestors. ∇

## 5. Cleaning the Pointing Graph

We investigate the $M^{\alpha}$-reachability of the elements not included in the $\alpha$-domain based on the given special decomposition. So, we will assume that a special decomposition of the set $S$ is given such that it satisfies the following conditions:

- $\{e_{q_1}, \ldots, e_{q_l}\} \subseteq S \backslash M^{\alpha}$,

- $\{M_{j_1}^{1-\alpha}, \ldots, M_{j_p}^{1-\alpha}\} \subseteq sM^{1-\alpha}$ is the set of main subsets in the (1-$\alpha$)-domain associated with the set $\{e_{q_1}, \ldots, e_{q_l}\}$.

By Proposition 4.5, if the elements $\{e_{q_1}, \ldots, e_{q_l}\}$ are $M^{\alpha}$-reachable, then the description of the replaceability procedure contains all ordered pairs that ensure their reachability.

During the replacement procedure, the elements of the set $\{e_{q_1}, \ldots, e_{q_l}\}$ are moved to the $\alpha$-domain as a result of the main replacement steps.

Therefore, if the replacement steps corresponding to the vertices of the graph did not lead to $M^{\alpha}$-reachability of all elements of the set $\{e_{q_1}, \ldots, e_{q_l}\}$, then some other elements were moved from the $\alpha$-domain to the (1-$\alpha$)-domain during the replacement procedure.

Below we will show that this is possible only in the following cases:

- the graph contains useless vertices that prevents some paths from reaching the final vertex, which means that corresponding sequence of replacement procedure is interrupted.

- there is an incompatible set of vertices in the graph (will be described later).

Therefore, we will describe procedures that will eliminate useless vertices and incompatible sets of vertices in a graph, if possible, so that as a result of these procedures, all elements located in the $\alpha$-domain of the initial special decomposition, will remain in the $\alpha$-domain of the resulting decomposition.

### 5.1. *The Removal Procedure.*

The procedure of successive removal of certain vertices and edges in the $M^{\alpha}$-pointing graph described below will be called the removal procedure.

The procedure starts by removing a specific vertex of the graph, let's denote it by $v$, and consists of the following steps:

(i) we remove the vertex $v$ and all edges outgoing from it.

- if, as a result of the removal of these edges, vertices with zero indegree appear, except for the main vertices, we mark them as generations of already removed vertex so that we can remove them later.

(ii) we sequentially examine the edges entering the vertex $v$.

Let all edges outgoing from some vertex $v_j$ and entering vertex $v$ be associated with the elements $e_{r_1}, \ldots, e_{r_l}$, respectively.

(ii.1) if all they are disjunctive edges, let them be the edges

$$(v_j, \vee e_{r_1}, v), \ldots, (v_j, \vee e_{r_l}, v),$$

such that for any $e \in \{e_{r_1}, \ldots, e_{r_l}\}$, there are other edges associated with the element $e$, then we remove all these edges and consider other edges incoming the vertex $v$, if there is one.

(ii.2) if for some element $e \in \{e_{r_1}, \ldots, e_{r_l}\}$, the edge $(v_j, \vee e, v)$ is the only disjunctive edge associated with $e$ (because the other disjunctive edges associated with $e$ have already been removed), we will call it the single-disjunctive edge associated with $e$, or it is a conjunctive edge, $(v_j, \& e, v)$, then:

- we remove all edges outgoing from the vertex $v_j$ and entering the vertex $v$,
- we mark the vertex $v_j$ as an ancestor of already removed vertex so that we can remove it later, and move on to considering other edges incoming the vertex $v$, if there is one.

(iii) to continue the procedure, we consider the marked vertices sequentially:

- if the vertex under consideration is marked as a generation, the point i) is applied to it.
- if the vertex under consideration is marked as an ancestor, we remove this vertex, and apply points i) and ii) to it.

The procedure ends if there are no more marked vertices.

In other words, if the edges $(v_{j_1}, v_i), \ldots, (v_{j_r}, v_i)$ and $(v_i, v_{k_1}), \ldots, (v_i, v_{k_s})$ are included in the graph and we remove the vertex $v_i$, then:

- all edges $(v_{j_1}, v_i), \ldots, (v_{j_r}, v_i)$ and $(v_i, v_{k_1}), \ldots, (v_i, v_{k_s})$ are removed,
- for any $v_j \in \{v_{j_1}, \ldots, v_{j_r}\}$, the vertex $v_j$ is removed only if $(v_j, v_i)$ is a conjunctive edge, or $(v_j, v_i)$ is the single-disjunctive edge associated with some element.
- for any $v_k \in \{v_{k_1}, \ldots, v_{k_s}\}$, the vertex $v_k$ is removed only if, as a result of removing the edges $(v_i, v_{k_1}), \ldots, (v_i, v_{k_s})$, the indegree of the vertex $v_k$ becomes zero.

*Definition* 5.1.1. A vertex in the pointing graph will be called removable if, as a result of applying the removal procedure to it, for any element $e \in \{e_{q_1}, \ldots, e_{q_l}\}$, some main vertex associated with $e$ is not removed.

5.1.2. A vertex in the pointing graph will be called not removable if, as a result of applying the removal procedure to it, for some element $e \in \{e_{q_1}, \ldots, e_{q_l}\}$, all main vertex associated with $e$ are removed.

5.2. *The Cleaning Procedure of The Graph.*

The procedure described below for eliminating the useless vertices in the $M^{\alpha}$-pointing graph will be called a cleaning procedure of the graph.

If a useless vertex is found in the graph, we apply the Removal Procedure to it.

If it is a removable vertex, then we conclude that the current useless vertex has been eliminated, and proceed to eliminate other useless vertices in the resulting graph, if there is any.

If we find a useless vertex that is not removable, we conclude that it is impossible to eliminate useless vertices in the graph, and the cleaning procedure is suspended.

*The graph* $G(e_{q_1}, \ldots, e_{q_l})$ *is called clean graph if it does not contain useless vertices and for any element* $e \in \{e_{q_1}, \ldots, e_{q_l}\}$ *it contains some main vertex associated with* $e$.

<u>*Remark*</u> 5.2.1. Recall that if $\{v_{i_1}, \ldots, v_{i_r}\}$ is the set of vertices of a graph $G(e_{q_1}, \ldots, e_{q_l})$, then the decomposition corresponding to these vertices is defined as $(i_1, \ldots, i_r)I(d_nS)$. That is, to get the corresponding decomposition, we permute the components of the ordered pairs

$$(M_{i_1}^{\alpha}, M_{i_1}^{1-\alpha}), \ldots, (M_{i_r}^{\alpha}, M_{i_r}^{1-\alpha})$$

in the initial decomposition. Therefore, removing a vertex in a graph means that the components of the ordered pair corresponding to that vertex will not be permuted.

*The graph obtained as a result of the cleaning procedure of the pointing graph* $G(e_{q_1}, \ldots, e_{q_l})$ *will also be called a clean graph corresponding to the replaceability procedure associated with the set* $\{e_{q_1}, \ldots, e_{q_l}\}$ *and will be denoted by* $CG(e_{q_1}, \ldots, e_{q_l})$.

For any main vertex $v_i$, the subgraph of the graph $CG(e_{q_1}, \ldots, e_{q_l})$ corresponding to the graph $G[v_i]$ will be denoted by $CG[v_i]$.

If the graph $G$ has no useless vertices, then $CG$ coincides with $G$.

<u>*Definition*</u> 5.3. Let a special decomposition $d_nS$ be given such that for some $\alpha \in \{0,1\}$,

$$\{e_{q_1}, \ldots, e_{q_l}\} = S \backslash M^{\alpha}.$$

5.3.1. An element $e \in S$ is called active in the $M^{\alpha}$-replaceability procedure associated with the set $\{e_{q_1}, \ldots, e_{q_l}\}$ if $e \in \{e_{q_1}, \ldots, e_{q_l}\}$, or the description of the replaceability procedure includes a step $(M_i^{\alpha}, M_i^{1-\alpha})$ such that $e \in M_i^{\alpha}$ and $M_i^{\alpha}$ is an $M^{\alpha}$-single subset with respect to the element $e$.

5.3.2. An element $e \in S$ will be called passive in the $M^{\alpha}$-replaceability procedure if it is not an active element.

5.3.3. The set $\{e_{q_1}, \ldots, e_{q_l}\}$ will be called a stable set with respect to the removal procedure started from some vertex of the $M^{\alpha}$-pointing graph $G(e_{q_1}, \ldots, e_{q_l})$, if as a result of the procedure, for every element $e \in \{e_{q_1}, \ldots, e_{q_l}\}$, some of the main vertices associated with $e$ are not removed.

5.3.4. The set $\{e_{q_1}, \ldots, e_{q_l}\}$ will be called a stable set with respect to the cleaning procedure of the $M^{\alpha}$-pointing graph $G(e_{q_1}, \ldots, e_{q_l})$, if as a result of the procedure all useless vertices in the graph are eliminated.

*Theorem* 5.4. Let $d_nS$ be a special decomposition such that $\{e_{q_1}, \dots, e_{q_l}\} \subseteq S \backslash M^\alpha$.

$\{v_{i_1}, \dots, v_{i_p}\}$ is the set of vertices of the $M^\alpha$-pointing graph $G(e_{q_1}, \dots, e_{q_l})$,

$\{v_{r_1}, \dots, v_{r_q}\}$ is the set of graph vertices obtained as a result of applying the removal procedure to some vertex $v_i \in \{v_{i_1}, \dots, v_{i_p}\}$.

If the set $\{e_{q_1}, \dots, e_{q_l}\}$ is stable with respect to the removal procedure started from the vertex $v_i$, then the set $(r_1, \dots, r_q)M^\alpha$ contains all active elements of the replaceability procedure associated with the set $\{e_{q_1}, \dots, e_{q_l}\}$.

*Proof.* Recall that $(r_1, \dots, r_q)M^\alpha$ is the set of elements included in the $\alpha$-components of the decomposition $(r_1, \dots, r_q)I(d_nS)$.

In fact, we need to prove that if the removal procedure is applied to some vertex, denote it by $v_i$, of the graph $G(e_{q_1}, \dots, e_{q_l})$, then the active elements of the replaceability procedure are not lose from the $\alpha$-domain. This will mean that if any active element is removed from the $\alpha$-domain of the decomposition $d_nS$ during the removal procedure, then it will be restored to the $\alpha$-domain of the decomposition $(r_1, \dots, r_q)I(d_nS)$ during the same procedure.

Note that if the vertex $v_i$ corresponds to the ordered pair $(M_i^\alpha, M_i^{1-\alpha})$ under the special decomposition $d_nS$, then also the ordered pair $(M_i^{1-\alpha}, M_i^\alpha)$ will correspond to $v_i$ under the special decomposition $(i_1, \dots, i_p)I(d_nS)$.

Assume that when the vertex $v_i$ is removed, some active element $e \in S$ is moved to the (1- $\alpha$)-domain. We consider the following cases:

(i) Let $v_i$ be a main vertex of the graph. This means that:

$$(e \in \{e_{q_1}, \dots, e_{q_l}\}) \,\&\, (e \in M_i^{1-\alpha}).$$

The ordered pair $(M_i^{1-\alpha}, M_i^\alpha)$ corresponds to the vertex $v_i$ in the special decomposition $(i_1, \dots, i_p)I(d_nS)$. Therefore, when the vertex $v_i$ is removed, the corresponding ordered pair takes its previous form, $(M_i^\alpha, M_i^{1-\alpha})$, that is, $e \in \{e_{q_1}, \dots, e_{q_l}\}$ moves back to the (1-$\alpha$)-domain. But since $e \in \{e_{q_1}, \dots, e_{q_l}\}$ and the set $\{e_{q_1}, \dots, e_{q_l}\}$ is stable with respect to the removal procedure started from the vertex $v_i$, there exists one more main vertex associated with the element $e$, which is not removed. Therefore, the element $e$ remains in the $\alpha$-domain of the decomposition $(r_1, \dots, r_q)I(d_nS)$.

(ii) Let $v_i$ not be the main vertex of the graph.

Since $v_i$ corresponds to the ordered pair $(M_i^\alpha, M_i^{1-\alpha})$ under the decomposition $d_nS$ and $e$ is an active element, this means that:

- the ordered pair $(M_i^\alpha, M_i^{1-\alpha})$ is included in the replaceability procedure and $e \in M_i^{1-\alpha}$,
- there is another ordered pair $(M_j^\alpha, M_j^{1-\alpha})$ included in the replaceability procedure such that $e \in M_j^\alpha$ and $M_j^\alpha$ is an $M^\alpha$-single subset with respect to the element $e$,
- the replacement step $(M_j^\alpha, M_j^{1-\alpha})$ leads to the step $(M_i^\alpha, M_i^{1-\alpha})$ to reach the element $e$.

So, the graph $G(e_{q_1}, \ldots, e_{q_l})$ contains the vertex $v_j$ and an edge $(v_j, v_i)$ associated with the element $e$. Recall that in the decomposition $(i_1, \ldots, i_p)I(d_nS)$ corresponding to the vertices of the graph $G(e_{q_1}, \ldots, e_{q_l})$, the vertices $v_j$ and $v_i$ correspond the ordered pairs $(M_j^{1-\alpha}, M_j^{\alpha})$ and $(M_i^{1-\alpha}, M_i^{\alpha})$, respectively. Removing the vertex $v_i$ means that the ordered pair $(M_i^{1-\alpha}, M_i^{\alpha})$ turns to ordered pair $(M_i^{\alpha}, M_i^{1-\alpha})$ so the element $e$ moves to the (1- $\alpha$)-domain.

According to the removal procedure, when removing the vertex $v_i$, we remove all edges entering $v_i$ and outgoing from it. As to the vertex $v_j$, we consider the following cases:

(i) let $(v_j, v_i)$ be a disjunctive edge associated with the element $e$ such that there is another edge $v_k$ associated with $e$. That is, the edges $(v_j, \vee e, v_i)$ and $(v_j, \vee e, v_k)$ are included in the graph. In this case we do not remove the vertex $v_j$, since the element $e$ will move to the $\alpha$-domain by permutation of ordered pair $(M_k^{\alpha}, M_k^{1-\alpha})$ corresponding to the vertex $v_k$.

(ii) let $(v_j, v_i)$ be a conjunctive edge or it is the single-disjunctive edge associated with the element $e$.

In this case we remove also the vertex $v_j$, which means that the corresponding ordered pair $(M_j^{1-\alpha}, M_j^{\alpha})$ returns to its previous form, that is, $(M_j^{\alpha}, M_j^{1-\alpha})$. So, with the subset $M_j^{\alpha}$ the element $e$ is returned to the $\alpha$-domain of the corresponding decomposition.

Obviously, by similar reasoning, we can conclude that no active element is lost from the $\alpha$-domain when other vertices are removed during this procedure. ∇

*Corollary* 5.4.1. Let $d_nS$ be a special decomposition such that $\{e_{q_1}, \ldots, e_{q_l}\} \subseteq S\backslash M^{\alpha}$.

$\{v_{r_1}, \ldots, v_{r_q}\}$ is the set of vertices of the cleaned graph $CG(e_{q_1}, \ldots, e_{q_l})$ obtained as a result of applying the cleaning procedure to the pointing graph $G(e_{q_1}, \ldots, e_{q_l})$.

If the set $\{e_{q_1}, \ldots, e_{q_l}\}$ is stable with respect to the cleaning procedure in the pointing graph $G(e_{q_1}, \ldots, e_{q_l})$, then the set $(r_1, \ldots, r_q)M^{\alpha}$ contains all active elements of the replaceability procedure under the decomposition $d_nS$.

*Proof.* Let the set $\{e_{q_1}, \ldots, e_{q_l}\}$ be stable with respect to the cleaning procedure.

The procedure for eliminating useless vertices consists in sequentially applying the removal procedure to these vertices. So, according to the Theorem 5.4, the set $(r_1, \ldots, r_q)M^{\alpha}$ contains all active elements of the replaceability procedure. ∇

*Corollary* 5.4.2. Let $d_nS$ be a special decomposition such that $\{e_{q_1}, \ldots, e_{q_l}\} = S\backslash M^{\alpha}$.

$CG(e_{q_1}, \ldots, e_{q_l})$ is the graph obtained as a result of applying the cleaning procedure to the pointing graph $G(e_{q_1}, \ldots, e_{q_l})$.

If the set $\{e_{q_1}, \ldots, e_{q_l}\}$ is stable with respect to the cleaning procedure of the graph, and all elements of the set $S$ are active in the replaceability procedure, then the $\alpha$-domain of the decomposition corresponding to the graph $CG(e_{q_1}, \ldots, e_{q_l})$ is a special covering for the set $S$.

*Proof.* Follows from the Theorem 5.4 and the Corollary 5.4.1. ∇

*Theorem* 5.5. Let $R$ be a description of the replaceability procedure associated with the set $\{e_{q_1}, \ldots, e_{q_l}\}$ under the decomposition $d_nS$, and $CG(e_{q_1},\ldots,e_{q_l})$ a clean graph with vertices $v_{i_1}, \ldots, v_{i_k}$ corresponding to the procedure $R$.

The vertex $v_i \in \{v_{i_1}, \ldots, v_{i_k}\}$ is removable in the graph $CG(e_{q_1},\ldots,e_{q_l})$ if and only if as a result of the replaceability procedure started from the ordered pair $(M_i^{1-\alpha}, M_i^{\alpha})$ under the decomposition $(i_1,\ldots,i_k)I(d_nS)$, the active elements of the replaceability procedure $R$ are not lost from the $\alpha$-domain.

*Proof.* Assume that $v_i$ is a removable vertex in the graph $CG(e_{q_1},\ldots,e_{q_l})$.

In this case, for any $e \in \{e_{q_1}, \ldots, e_{q_l}\}$, some main vertex associated with $e$ is not removed when applying the removal procedure to the vertex $v_i$. Let for given element $e$, the main vertex $v_j$ is not removed.

This means that any path leading from the vertex $v_j$ to the vertex $v_i$ contains a pair of adjacency vertices, let for particular path them be $v_{k_1}$ and $v_{k_2}$, such that the following holds:

i) all edges outgoing from $v_{k_1}$ to $v_{k_2}$ are disjunctive,

ii) if for some element, $e_r$, there exists an edge $(v_{k_1}, \vee e_r, v_{k_2})$ then for some vertex $v$ the graph contains the edge $(v_{k_1}, \vee e_r, v)$, which is not included in any path leading from the main vertex $v_j$ to the vertex $v_i$.

Otherwise, when the vertex $v_{k_2}$ is removed, the vertex $v_{k_1}$ will also be removed, which will lead to the removal of the vertex $v_j$.

Thus, if the path contains adjacency vertices $v_{k_1}$ and $v_{k_2}$ satisfying conditions i) and ii), then the removal procedure applied to the vertex $v_i$ will be interrupted when the vertex $v_{k_1}$ is reached, moving in the opposite direction along the considered path. So, the procedure does not reach to the vertex $v_j$. The result will be the same for other paths going from vertex $v_j$ to the vertex $v_i$.

Let us prove that the active elements of the replaceability procedure $R$ are not lost from the $\alpha$-domain, as a result of the replacement procedure started from the step $(M_i^{1-\alpha}, M_i^{\alpha})$ in the decomposition $(i_1,\ldots,i_k)I(d_nS)$.

Recall that any path leading from the main vertex $v_j$ to the vertex $v_i$ contains a pair of adjacency vertices satisfying the conditions i) and ii). Let each of the following pairs

$$(v_{r_1}, v_{l_1}), (v_{r_2}, v_{l_2}), \ldots, (v_{r_s}, v_{l_s})$$

be the last pair of such vertices on different paths, respectively.

This means that for any $v_{r_k} \in \{v_{r_1}, \ldots, v_{r_s}\}$, all edges on the path leading from $v_{r_k}$ to $v_i$ are conjunctive edges, except for the edge outgoing from $v_{r_k}$, which is a disjunctive edge. Let them be the following edges:

$$(v_{r_k}, \vee e, v_{l_k}), (v_{l_k}, \&e_1, v_{s_1}), \ldots, (v_{s_{p-1}}, \&e_p, v_{s_p}), (v_{s_p}, \&e_i, v_i).$$

The vertices $v_{r_k}$, $v_{l_k}$, $v_{s_1}$, . . ., $v_{s_p}$, $v_i$ are included the in graph $CG(e_{q_1},\dots,e_{q_l})$, so the ordered pairs,

$$(M^{\alpha}_{r_k}, M^{1-\alpha}_{r_k}), (M^{\alpha}_{l_k}, M^{1-\alpha}_{l_k}), (M^{\alpha}_{s_1}, M^{1-\alpha}_{s_1}), \dots, (M^{\alpha}_{s_p}, M^{1-\alpha}_{s_p}), (M^{\alpha}_{i}, M^{1-\alpha}_{i})$$

will correspond to them in the decomposition $d_nS$, and $e$, $e_1$, . . ., $e_p$, $e_i$ are the active elements.

Since the disjunctive edge $(v_{r_k}, \vee e, v_{l_k})$ is included in this path then the graph contains another disjunctive edge associated with $e$, which is not included in any path leading from $v_j$ to $v_i$. Let it be the edge $(v_{r_k}, \vee e, v_q)$ for some vertex $v_q$.

This means that in the decomposition $d_nS$ the following holds:

- $e \in M^{\alpha}_{r_k}$ and $M^{\alpha}_{r_k}$ is an $M^{\alpha}$-single subset with respect to $e$,
- $(e \in M^{1-\alpha}_{l_k})$ & $(e \in M^{1-\alpha}_{q})$,
- $e_1 \in M^{\alpha}_{l_k}$ and $M^{\alpha}_{l_k}$ is an $M^{\alpha}$-single subset with respect to $e_1$,
- $e_1 \in M^{1-\alpha}_{s_1}$ and $M^{1-\alpha}_{s_1}$ is an $M^{1-\alpha}$-single subset with respect to $e_1$,
- $e_2 \in M^{\alpha}_{s_1}$ and $M^{\alpha}_{s_1}$ is an $M^{\alpha}$-single subset with respect to $e_2$,
- $e_2 \in M^{1-\alpha}_{s_2}$ and $M^{1-\alpha}_{s_2}$ is an $M^{1-\alpha}$-single subset with respect to $e_1$,

. . .

- $e_p \in M^{\alpha}_{s_{p-1}}$ and $M^{\alpha}_{s_{p-1}}$ is an $M^{\alpha}$-single subset with respect to $e_{p-1}$,
- $e_p \in M^{1-\alpha}_{s_p}$ and $M^{1-\alpha}_{s_p}$ is an $M^{\alpha}$-single subset with respect to $e_p$,
- $e_i \in M^{\alpha}_{s_p}$ and $M^{\alpha}_{s_p}$ is an $M^{\alpha}$-single subset with respect to $e_i$,
- $e_i \in M^{1-\alpha}_{i}$ and $M^{1-\alpha}_{i}$ is an $M^{1-\alpha}$-single subset with respect to $e_i$.

The corresponding ordered pairs in the decomposition $(i_1,\dots,i_k)I(d_nS)$ will look like

$$(M^{1-\alpha}_{r_k}, M^{\alpha}_{r_k}), (M^{1-\alpha}_{l_k}, M^{\alpha}_{l_k}), (M^{1-\alpha}_{s_1}, M^{\alpha}_{s_1}), \dots, (M^{1-\alpha}_{s_p}, M^{\alpha}_{s_p}), (M^{1-\alpha}_{i}, M^{\alpha}_{i}).$$

Note that since the subsets

$$M^{\alpha}_{l_k}, M^{\alpha}_{s_1}, \dots, M^{\alpha}_{s_p}$$

are $M^{\alpha}$-single with respect to the elements $e_1$, . . ., $e_p$, $e_i$ respectively, then these elements cannot be included in the $\alpha$-components of the ordered pairs not included in the description of the replaceability procedure associated with the set $\{e_{q_1}, \dots, e_{q_l}\}$ under the decomposition $d_nS$.

Hence, the subsets

$$M^{1-\alpha}_{s_1}, \dots, M^{1-\alpha}_{s_p}, M^{1-\alpha}_{i}$$

are also $M^{\alpha}$-single with respect to the elements $e_1$, . . ., $e_p$, $e_i$ respectively, under the decomposition $(i_1, \dots, i_k)I(d_nS)$. The subset $M^{1-\alpha}_{l_k}$ is not $M^{\alpha}$-single with respect to $e$ in the decomposition $(i_1, \dots, i_k)I(d_nS)$, since $e$ is also included in $M^{1-\alpha}_{q}$.

This means that the following set of ordered pairs,

$$\{(M^{1-\alpha}_{i}, M^{\alpha}_{i}), (M^{1-\alpha}_{s_p}, M^{\alpha}_{s_p}), \dots, (M^{1-\alpha}_{s_1}, M^{\alpha}_{s_1}), (M^{1-\alpha}_{l_k}, M^{\alpha}_{l_k})\},$$

of the decomposition $(i_1,\dots,i_k)I(d_nS)$, is part of replaceability procedure started from the replacement step $(M^{1-\alpha}_{i}, M^{\alpha}_{i})$.

In addition:

- the ordered pairs

$$(M_i^{1-\alpha}, M_i^{\alpha}), (M_{s_p}^{1-\alpha}, M_{s_p}^{\alpha}), \ldots, (M_{s_1}^{1-\alpha}, M_{s_1}^{\alpha}), (M_{l_k}^{1-\alpha}, M_{l_k}^{\alpha})$$

are obligatory replacement steps in the replaceability procedure started from the replacement step $(M_i^{1-\alpha}, M_i^{\alpha})$ in the decomposition $(i_1, \ldots, i_k)I(d_nS)$,

- $(M_{l_k}^{1-\alpha}, M_{l_k}^{\alpha})$ is a final replacement step for the replacement procedure started from the step $(M_i^{1-\alpha}, M_i^{\alpha})$ in the decomposition $(i_1, \ldots, i_k)I(d_nS)$,

- $(M_{r_k}^{1-\alpha}, M_{r_k}^{\alpha})$ do not enter the replaceability procedure started from the replacement step $(M_i^{1-\alpha}, M_i^{\alpha})$ in the decomposition $(i_1, \ldots, i_k)I(d_nS)$.

Therefore, the procedure along the considered path will be suspended when the step $(M_{l_k}^{1-\alpha}, M_{l_k}^{\alpha})$ is reached. As a result, the elements $e_1, \ldots, e_p, e_i$, which are active also during this procedure will not be lost from the $\alpha$-domain. Since the step $(M_q^{1-\alpha}, M_q^{\alpha})$ is not included in the replaceability procedure, the element $e$ remains in the $\alpha$-domain.

Now suppose that as a result of the replacement procedure in the special decomposition $(i_1, \ldots, i_k)I(d_nS)$ started from the step $(M_i^{1-\alpha}, M_i^{\alpha})$, the active elements of the procedure $R$ are not lost from the $\alpha$-domain.

We will prove that the vertex $v_i$ is removable in the graph $CG(e_{q_1}, \ldots, e_{q_l})$.

Let's denote by $R_1$ the description of the replaceability procedure started from the ordered pair $(M_i^{1-\alpha}, M_i^{\alpha})$:

$$R_1 = \{(M_i^{1-\alpha}, M_i^{\alpha}), (M_{k_1}^{1-\alpha}, M_{k_1}^{\alpha}), \ldots, (M_{k_r}^{1-\alpha}, M_{k_r}^{\alpha})\}$$

Assume that $v_i$ is a non-removable vertex in the graph $CG(e_{q_1}, \ldots, e_{q_l})$ and discuss the following possible cases:

1) $v_i$ is the only main vertex in the graph associated with some element $e$.

This means that the ordered pair $(M_i^{\alpha}, M_i^{1-\alpha})$ is the only main replacement step associated with $e$. But then as a result of the replacement procedure started from $(M_i^{1-\alpha}, M_i^{\alpha})$ under the decomposition $(i_1, \ldots, i_k)I(d_nS)$, the element $e$ will be lost from the $\alpha$-domain, which will contradict the activity condition for the element $e$ in $d_nS$. So, $v_i$ cannot be the only main vertex associated with the element $e$.

2) for some element $e \in \{e_{q_1}, \ldots, e_{q_l}\}$, from any main vertex associated with $e$, a path leads to the vertex $v_i$, such that no adjacent vertices in it satisfy conditions i) and ii).

Let $v_j$ be a main vertex associated with the element $e$, and

$$v_j, (v_j, v_{l_1}), v_{l_1}, \ldots, v_{l_k}, (v_{l_k}, v_i), v_i$$

be a path in which there are no adjacent vertices that satisfy conditions i) and ii)).

It is easy to see that it suffices to consider the case when all edges of this path are conjunctive. That is, the path looks like

$$(v_j, \&e_{p_1}, v_{l_1}), (v_{l_1}, \&e_{p_2}, v_{l_2}), \ldots, (v_{l_{k-1}}, \&e_{p_k}, v_{l_k})\ (v_{l_k}, \&e_i, v_i),$$

where the elements $e_{p_1},\dots,e_{p_k}$, $e_i$ are not included in $\alpha$-component of any ordered pair not included in the description of the replaceability procedure $R$.

This means that the corresponding replaceability steps under the decomposition $d_nS$ are obligatory steps and the following holds:

- $e \notin M^\alpha$ & $e \in M_j^{1-\alpha}$
- $e_{p_1} \in M_j^\alpha$ and $M_j^\alpha$ is an $M^\alpha$-single subset with respect to $e_{p_1}$,
- $e_{p_1} \in M_{l_1}^{1-\alpha}$ and $M_{l_1}^{1-\alpha}$ is $M^{1-\alpha}$-single subset with respect to $e_{p_1}$,

. . .

- $e_{p_k} \in M_{l_{k-1}}^\alpha$ and $M_{l_{k-1}}^\alpha$ is an $M^\alpha$-single subset with respect to $e_{p_k}$,
- $e_{p_k} \in M_{l_k}^{1-\alpha}$ and $M_{l_k}^{1-\alpha}$ is an $M^\alpha$-single subset with respect to $e_{p_k}$,
- $e_i \in M_{l_k}^\alpha$ and $M_{l_k}^\alpha$ is an $M^\alpha$-single subset with respect to $e_i$,
- $e_i \in M_i^{1-\alpha}$ and $M_i^{1-\alpha}$ is an $M^{1-\alpha}$-single subset with respect to $e_i$.

Evidently, by exploring another path leading from another main vertex associated with the element $e$ to the vertex $v_i$, we get the similar result. But then as a result of the replacement procedure of $R_1$, the element $e$ will be lost from the $\alpha$-domain, which again contradicts the condition that the element $e$ is active in $d_nS$. So, $v_i$ cannot be a non-removable vertex.

Thus, if $(M_i^{1-\alpha}, M_i^\alpha)$ is an ordered pair in the special decomposition

$$(i_1,\dots,i_k)I(d_nS),$$

and as a result of the replacement procedure started from it, the active elements of the replacement procedure $R$ are not lost from the $\alpha$-domain, then the corresponding vertex $v_i$ is removable. ∇

The Theorem 5.5 gives an important possibility:

To restore the lost element in the $\alpha$-domain, we find a removable vertex associated with this element and apply the removal procedure to it. As a result of the removal procedure started from this vertex, the lost element goes into the $\alpha$-domain of the obtained decomposition.

*Lemma* 5.6. Let $d_nS$ be a special decomposition such that $\{e_{q_1}, \dots, e_{q_l}\} \subseteq S\backslash M^\alpha$.

If $\{v_{j_1}, \dots, v_{j_l}\}$ is the set of main vertices of some clean graph $CG(e_{q_1}, \dots, e_{q_l})$, obtained as a result of the cleaning procedure in the graph $G(e_{q_1}, \dots, e_{q_l})$, then:

(i) for any $v \in \{v_{j_1}, \dots, v_{j_l}\}$, $CG[v]$ is a connected graph, and there are paths from the vertex $v$ to any of the vertices of the graph $CG[v]$.

(ii) for any non-final vertex $v_j \in CG[v]$ corresponding to an ordered pair $(M_j^\alpha, M_j^{1-\alpha})$, if the subset $M_j^\alpha$ is $M^\alpha$-single with respect to some element $e \in S$, then the graph $CG[v]$ contains a vertex $v_k$ associated with $e$ such that an edge $(v_j, v_k)$ is included in $CG[v]$.

*Proof.* i) By Proposition 4.7, the graph $G[v]$ is a connected graph, and there are paths from the vertex $v$ to any other vertices of the graph $G[v]$. If a disconnected part appears when some edges and vertices are removed during the cleaning procedure, then this part is removed during the same procedure. So, the point (i) is true.

(ii) Suppose that the vertex $v_j \in CG[v]$ corresponds to an ordered pair $(M_j^{\alpha}, M_j^{1-\alpha})$ such that $M_j^{\alpha}$ is an $M^{\alpha}$-single subset with respect to some element $e \in S$. This means that $v_j$ cannot be a useless vertex because otherwise would be removed during the cleaning procedure. Therefore, the graph $G[v]$ contains some vertex $v_k$ and an edge $(v_j, v_k)$ associated with the element $e$.

Consider the following cases:

1) let $(v_j, v_k)$ be a conjunctive edge $(v_j, \&e, v_k)$.

Since the vertex $v_j$ is not removed during the removal procedure, then the edge $(v_j, \&e, v_k)$ is also not removed because, as described in 5.1, when this edge is removed, $v_j$ will also be removed. So, $(v_j, \&e, v_k) \in CG[v]$.

2) let $(v_j, v_k)$ is a disjunctive edge $(v_j, \vee e, v_k)$.

In this case the graph $G[v]$ contains other disjunctive edges going from the vertex $v_j$ to other vertices and associated with the element $e$.

Obviously, since the vertex $v_j$ is not removed during the cleaning procedure, then at least one of these edges is also not removed, so, $(v_j, \vee e, v_k) \in CG[v]$, for some vertex $v_k \in CG[v]$. ∇

Let $u$ and $v$ be different useless vertices in the pointing graph $G$ associated with the set $\{e_{q_1}, \ldots, e_{q_l}\}$. We will use the following notation:

$(u)G$ is the graph obtained as a result of applying the removal procedure to the vertex $u$ in the graph $G$.

$(v)G$ is the graph obtained as a result of applying the removal procedure to the vertex $v$ in the graph $G$.

*Lemma* 5.7. (i) If the vertex $u$ is not removable in the graph $G$, and the vertex $v$ is removable in $G$, then $u$ also cannot be removable in the graph $(v)G$.

(ii) If each of the vertices $u$ and $v$ are removable in the graph $G$, then $v$ is removable in the graph $(u)G$ if and only if $u$ is removable in the graph $(v)G$.

*Proof.* (i) Since the vertex $v$ is removable in the graph $G$, then for any $e \in \{e_{q_1}, \ldots, e_{q_l}\}$ some main vertex associated with $e$ is not removed when applying the removal procedure to the vertex $v$.

Also, since the vertex $u$ is not removable, all main vertices associated with some element $e$ are removed when applying the removal procedure to $u$. Let them be the vertices $v_{i_1}, \ldots, v_{i_k}$.

We discuss the following cases:

(1) all edges and vertices included in the removal procedure to eliminate the vertex $u$ are included in the graph $(v)G$.

Obviously, in this case $u$ cannot be eliminated also in the graph $(v)G$.

(2) some vertices and edges that are removed during the procedure for eliminating the vertex $u$ are also removed during the procedure for eliminating the vertex $v$. We will refer to them as common vertices and edges for these procedures.

We show that the following statements hold:

(2.1) none of these common vertices coincides with the vertex $u$.

This is obvious, since otherwise the vertex $v$ would not have been eliminated.

(2.2) the absence of common vertices in the graph $(v)G$, which were removed as a result of applying the removal procedure to the vertex $v$, does not contribute to the preservation of the main vertices $v_{i_1}$, . . ., $v_{i_k}$, when the cleaning procedure is applying to the vertex $u$ in the graph $(v)G$.

During the cleaning procedure, any vertex of the graph is removed either due to the removal of another vertex or if the removal procedure is applied to it.

According to the description of the removal procedure, if some vertex is removed during both separate procedures started from the vertex $u$ and started from the vertex $v$ in the same graph, then the actions after the removal of this vertex are the same in both procedures. This means that if the removal of some vertices leads to the removal of the main vertex, then it does not depend on the procedure in which this occurs.

Thus, the removal of some vertices can only contribute to the removal of some main vertex, not its preservation.

In addition, the vertex $u$ is not removed during the removal procedure started from the vertex $v$. So, if the removal procedure is applied to the vertex $u$ in the graph $(v)G$, all edges and vertices that are removed during the removal procedure started from the vertex $u$ in the graph $G$ are removed, except for common edges and vertices that have already been removed. This means that the corresponding main vertices are also removed.

Therefore, the vertex $u$ cannot be eliminated in the graph $(v)G$.

(ii) Suppose that under the given conditions the vertex $u$ is eliminated in the graph $(v)G$, and the vertex $v$ is not eliminated in the graph $(u)G$.

This means that for some element $e \in \{e_{q_1}, . . ., e_{q_l}\}$, as a result of applying the removal procedure to the vertex $v$ in the graph $(u)G$, all the main vertices associated with $e$ are removed.

According to Proposition 4.7, for any vertex of the graph, there is a path passing from some main vertex to this vertex. Removing some vertex included in this path, we remove vertices, moving both in the direction of this path and in the opposite direction of this path.

Let for some vertices $v_p$, $v_q$ the edge $(v_p, v_q)$ be included in the path passing from some main vertex to the vertex $v_q$. Then, according to the removal procedure, the removal of the vertex $v_q$ leads to the removal of the vertex $v_p$ if $(v_p, v_q)$ is a conjunctive edge, or if $(v_p, v_q)$ is the single-disjunctive edge associated with some element $e$.

Suppose that, removing vertices, we move in the opposite direction along some path.

If a disjunctive edge associated with some element is encountered, and there are other disjunctive edges associated with the same element then, removing this edge, the procedure in this direction is suspended.

Therefore, since both the vertices $u$ and $v$ separately can be eliminated in the graph $G$, and $u$ is not eliminated in the graph $(v)G$, then both separate procedures for eliminating the vertices $u$ and $v$, include some common vertices, $v_{p_1}, \ldots, v_{p_r}$ such that:

- $v_{p_1}, \ldots, v_{p_r}$ are not removed during none of the separate procedures for eliminating the vertices $u$ and $v$ in the graph $G$,

- $v_{p_1}, \ldots, v_{p_r}$ are removed when performing the removal procedure for elimination the vertex $v$ in the graph $(u)G$.

- the removal of these vertices leads to the removal of some main vertices.

It is easy to see that this can only take place if each of these vertices has outgoing disjunctive edges associated with the same element, so that some of these disjunctive edges are removed during the elimination procedure for the vertex $u$ in the graph $G$, and others - during the elimination procedure for the vertex $v$ in the graph $G$.

Obviously, if these procedures are performed separately, none of these vertices will be removed. It is also obvious that as a result of sequential application of the removal procedure to vertex $u$, and then to vertex $v$, the vertices $v_{p_1}, \ldots, v_{p_r}$ will be removed. So, the corresponding main vertices will be removed.

But the vertices $v_{p_1}, \ldots, v_{p_r}$ will be removed also as a result of successive application of the removal procedures in opposite order that is, to the vertex $v$, and then to the vertex $u$, which means that the vertex $u$ is not eliminated in the graph $(v)G$. We get a contradiction. So, the point (ii) is true. ∇

*Corollary* 5.7.1. If different useless vertices are included in the pointing graph $G$, then the result of successive removal procedures to eliminate all these vertices does not depend on the order in which they are considered.

*Proof.* Let $v_{i_1}, \ldots, v_k$ be all useless vertices in the graph $G(e_{q_1}, \ldots, e_{q_l})$.

If any of these vertices is not removable in the graph $G(e_{q_1}, \ldots, e_{q_l})$, then according to the Lemma 5.7.(i), it will be not removable also in the graph $(v)G(e_{q_1}, \ldots, e_{q_l})$ for any vertex

$$v \in \{v_{i_1}, \ldots, v_k\}$$

which means it is impossible to eliminate useless vertices in the graph $G(e_{q_1}, \ldots, e_{q_l})$.

According to the Lemma 5.7.(ii), it is easy to see that if each of the vertices $v_{i_1}, \ldots, v_k$ is removable in $G(e_{q_1}, \ldots, e_{q_l})$, then the result of successive procedures for eliminating all these vertices is the same, when applying the procedures in any sequence. ∇

## 6. Incompatible Sets of Vertices in a Graph

We assume that a special decomposition $d_nS$ is given such that

- $\{e_{q_1}, \ldots, e_{q_l}\} \subseteq S \backslash M^{\alpha}$,

- $\{e_{q_1}, \ldots, e_{q_l}\}$ is a stable set of elements with respect to the cleaning procedure in the graph $G(e_{q_1}, \ldots, e_{q_l})$.

According to the Corollary 5.7.1, if the pointing graph $G$ includes different useless vertices, then the result of the cleaning procedure does not depend on the order in which the removal procedure is successively applied to each of them.

If the set $\{e_{q_1}, \ldots, e_{q_l}\}$ is not stable, with respect to the cleaning procedure in the graph, then all the main vertices associated with some element $e \in \{e_{q_1}, \ldots, e_{q_l}\}$ will be removed, which means that under the given decomposition there is no special covering for the set $S$.

On the other hand, according to Theorem 5.4, the stability of the set $\{e_{q_1}, \ldots, e_{q_l}\}$, as a result of the cleaning procedure in the graph, guarantees the presence of only active elements in the $\alpha$-domain. This means that some passive elements can move to the (1-$\alpha$)-region of the corresponding decomposition.

*Definition* 6.1. Let the set

$$R = \{(M_{r_1}^{\alpha},\ M_{r_1}^{1-\alpha}), \ldots, (M_{r_q}^{\alpha}, M_{r_q}^{1-\alpha})\}$$

be a description of the replaceability procedure associated with the set $\{e_{q_1}, \ldots, e_{q_l}\}$ under some special decomposition, and $G(e_{q_1}, \ldots, e_{q_l})$ corresponding pointing graph.

The set of ordered pairs

$$R_C = \{(M_{c_1}^{\alpha}, M_{c_1}^{1-\alpha}), \ldots, (M_{c_p}^{\alpha}, M_{c_p}^{1-\alpha})\}$$

correspond to the set of vertices of the clean graph $CG(e_{q_1}, \ldots, e_{q_l})$ obtained as a result of cleaning procedure in the graph $G(e_{q_1}, \ldots, e_{q_l})$.

The set of ordered pairs

$$R_I = \{(M_{i_1}^{\alpha}, M_{i_1}^{1-\alpha}), \ldots, (M_{i_k}^{\alpha}, M_{i_k}^{1-\alpha})\} \subseteq R_C$$

will be called an incompatible set with respect to some element $e \in S$, if the following conditions are satisfied:

- $k > 1$,

- $\forall i \in \{i_1, \ldots, i_k\}\ (e \in M_i^{\alpha})$,

- $e \notin (c_1, \ldots, c_p)\ M^{\alpha}$.

Accordingly, the corresponding set of vertices $\{v_{i_1}, \ldots, v_k\}$ in the clean graph will be called an incompatible set of vertices in the graph.

It is important to note, that:

- the incompatibility is defined only for the vertices of the clean graph,

- there can be several incompatible sets of vertices in the graph,

- different sets of vertices cannot be incompatible with respect to same element,

- a set of vertices may be incompatible with respect to different elements.

*Lemma* 6.2. Let a special decomposition of the set $S$ be given such that

$$\{e_{q_1}, \ldots, e_{q_l}\} \subseteq S \backslash M^{\alpha},$$

$R = \{(M_{r_1}^{\alpha}, M_{r_1}^{1-\alpha}), \ldots, (M_{r_q}^{\alpha}, M_{r_q}^{1-\alpha})\}$ is the description of a replaceability procedure associated with the set $\{e_{q_1}, \ldots, e_{q_l}\}$, and $G(e_{q_1}, \ldots, e_{q_l})$ is the corresponding pointing graph.

$R_C = \{(M_{c_1}^{\alpha}, M_{c_1}^{1-\alpha}), \ldots, (M_{c_p}^{\alpha}, M_{c_p}^{1-\alpha})\}$ is the set of ordered pairs corresponding to the set of vertices of the clean graph $CG(e_{q_1}, \ldots, e_{q_l})$.

If $\{e_{q_1}, \ldots, e_{q_l}\}$ is a stable set with respect to the cleaning procedure, then:

The set of ordered pairs

$$R_I = \{(M_{i_1}^{\alpha}, M_{i_1}^{1-\alpha}), \ldots, (M_{i_k}^{\alpha}, M_{i_k}^{1-\alpha})\} \subseteq R_c$$

will be incompatible with respect to some element $e \in S$, if and only if one of the following statements is true:

(i) $[k > 1]$ & $[\forall i \in \{i_1, \ldots, i_k\}(e \in M_i^{\alpha})]$ & $[e \notin (M^{\alpha} \setminus \cup_{j=1}^{k} M_{i_j}^{\alpha})]$ & $[e \notin \cup_{j=1}^{q} M_{r_j}^{1-\alpha}]$.

(ii) if the set of ordered pairs

$$\{(M_{d_1}^{\alpha}, M_{d_1}^{1-\alpha}), \ldots, (M_{d_r}^{\alpha}, M_{d_r}^{1-\alpha})\} \subset R$$

correspond to the vertices of the graph that are removed during some removal procedure in the graph, then $R_I$ can be represented as

$$R_I = \{(M_{l_1}^{\alpha}, M_{l_1}^{1-\alpha}), \ldots, (M_{l_r}^{\alpha}, M_{l_r}^{1-\alpha}), (M_{s_1}^{\alpha}, M_{s_1}^{1-\alpha}), \ldots, (M_{s_q}^{\alpha}, M_{s_q}^{1-\alpha})\}$$

such that the following holds for some $r > 1$ and $t \geq 1$:

- $[\forall l \in \{l_1, \ldots, l_r\}$ $(e \in M_l^{\alpha})]$ & $[\{M_{l_1}^{\alpha}, \ldots, M_{l_r}^{\alpha}\} \subseteq \{M_{r_1}^{\alpha}, \ldots, M_{r_q}^{\alpha}\} \setminus \{M_{d_1}^{\alpha}, \ldots, M_{d_r}^{\alpha}\}]$,
- $[\forall s \in \{s_1, \ldots, s_q\}$ $(e \in M_s^{1-\alpha})]$ & $[\{M_{s_1}^{1-\alpha}, \ldots, M_{s_t}^{1-\alpha}\} \subseteq \{M_{d_1}^{1-\alpha}, \ldots, M_{d_r}^{1-\alpha}\}]$.

*Proof.* One side of the lemma is obvious. It is easy to see that if there is a set of ordered pairs

$$\{(M_{i_1}^{\alpha}, M_{i_1}^{1-\alpha}), \ldots, (M_{i_k}^{\alpha}, M_{i_k}^{1-\alpha})\} \subseteq R$$

and an element $e \in S$ such that one of the statements (i) and (ii) holds, then this set is incompatible with respect to $e$.

Other side: Assume that the set of ordered pairs

$$R_I = \{(M_{i_1}^{\alpha}, M_{i_1}^{1-\alpha}), \ldots, (M_{i_k}^{\alpha}, M_{i_k}^{1-\alpha})\} \subseteq R_C$$

is incompatible with respect to some element $e \in S$. That is, the following conditions are satisfied:

- $k > 1$,
- $\forall i \in \{i_1, \ldots, i_k\}$ $(e \in M_i^{\alpha})$,
- $e \notin (c_1, \ldots, c_p)$ $M^{\alpha}$.

This means that as a result of replacement and cleaning procedures the element $e$, that is not included in the set $\{e_{q_1}, \ldots, e_{q_l}\}$, moves from the $\alpha$-domain of the given decomposition to the (1-$\alpha$)-domain of the decomposition corresponding to the vertices of the graph

$$CG(e_{q_1}, \ldots, e_{q_l}).$$

So, the $\alpha$-domain of this decomposition does not cover the set $S$. According to Theorem 5.5, the element $e$ will be passive.

We will prove that one of the statements (i) and (ii) holds.

The elements of the set $\{e_{q_1}, \ldots, e_{q_l}\}$ go to the $\alpha$-domain as a result of the replacement procedure. Since the set $\{e_{q_1}, \ldots, e_{q_l}\}$ is stable with respect to the cleaning procedure in the graph $G(e_{q_1}, \ldots, e_{q_l})$, then these elements will also enter in the $\alpha$-domain of the decomposition corresponding to the vertices of the clean graph $CG(e_{q_1}, \ldots, e_{q_l})$.

Let's discuss when and how the element $e$ moves to the (1-$\alpha$)-domain.

To obtain a clean graph based on the given decomposition, we do the following:

- find all ordered pairs that make up the replacement procedure and construct the corresponding pointing graph,
- apply the removal procedure to each useless vertex in order to eliminate them, if there are any.

We obtain the decomposition corresponding to the vertices of the resulting graph by permuting the components of the ordered pairs corresponding to the vertices of the graph.

Thus, we perform the replacement and the removal procedure. So, any element can move to the (1-$\alpha$)-domain during one of these procedures.

Consider the following cases:

1) The element $e$ moves to the (1-$\alpha$)-domain during the replacement procedure.

Since $e$ is a passive element and is not restored in the $\alpha$-domain, the subset containing this element is not $M^\alpha$-single with respect to $e$. Otherwise, either $e$ would return to the $\alpha$-domain by permuting the components of another pair, according to the graph construction procedure, or the corresponding vertex of the graph would become a useless vertex, whose removal will return $e$ to the $\alpha$-domain.

Thus, we have the following:

$e \in M^\alpha$ but no subset in the $\alpha$-domain is $M^\alpha$-single with respect to $e$. This means that the replaceability procedure contains the ordered pairs,

$$\{(M_{i_1}^{\alpha}, M_{i_1}^{1-\alpha}), \ldots, (M_{i_k}^{\alpha}, M_{i_k}^{1-\alpha})\},$$

for some $k > 1$, such that $e$ is included in all their $\alpha$-components, that is,

$\forall i \in \{i_1, \ldots, i_k\}\, (e \in M_i^\alpha)$ and $e \notin (M^\alpha \setminus \bigcup_{j=1}^{k} M_{i_i}^{\alpha})$.

In addition, $e \notin \bigcup_{j=1}^{q} M_{r_j}^{1-\alpha}$ since as a result of replacement the $\alpha$-domain loses the element $e$. Thus, the statement (i) holds.

2) The element $e$ moves to the (1-$\alpha$)-domain during the removal procedure.

Suppose that the ordered pairs

$$\{(M_{d_1}^{\alpha}, M_{d_1}^{1-\alpha}), \ldots, (M_{d_p}^{\alpha}, M_{d_p}^{1-\alpha})\},$$

correspond to the vertices $v_{d_1}, \ldots, v_{d_r}$ that are removed during the cleaning procedure in the graph $G(e_{q_1}, \ldots, e_{q_l})$.

Recall that, as a result of permutation, these ordered pairs look like

$$\{(M_{d_1}^{1-\alpha}, M_{d_1}^{\alpha}), \ldots, (M_{d_p}^{1-\alpha}, M_{d_p}^{\alpha})\},$$

and also, removing the corresponding vertices restores the former order of their components.

By the condition, as a result of removing these vertices, the element $e$ goes to the (1-$\alpha$)-domain. This means that after the replacement procedure, $e$ occurs in the $\alpha$-domain only in some of the subsets $M_{d_1}^{1-\alpha}, \ldots, M_{d_p}^{1-\alpha}$. Let them be the subsets

$$\{M_{s_1}^{1-\alpha}, \ldots, M_{s_t}^{1-\alpha}\} \subseteq \{M_{d_1}^{1-\alpha}, \ldots, M_{d_p}^{1-\alpha}\}.$$

It is easy to see that $t \geq 1$. These subsets are not included in the set of main subsets of the original decomposition as $e \notin \{e_{q_1}, \ldots, e_{q_l}\}$. Therefore, also $e \in M^{\alpha}$. Let $M_{l_1}^{\alpha}, \ldots, M_{l_r}^{\alpha}$ be the subsets in the $\alpha$-domain of the initial decomposition containing the element $e$. It is easy to notice that

$$\{M_{l_1}^{\alpha}, \ldots, M_{l_r}^{\alpha}\} \subseteq \{M_{r_1}^{\alpha}, \ldots, M_{r_q}^{\alpha}\} \setminus \{M_{d_1}^{\alpha}, \ldots, M_{d_p}^{\alpha}\}.$$

Let's show that none of these subsets is $M^{\alpha}$-single with respect to $e$, that is, $r > 1$.

If $r$ =1, then the subset $M_{l_1}^{\alpha}$ is $M^{\alpha}$-single and so, the replacement step $(M_l^{\alpha}, M_l^{1-\alpha})$ will lead to the replacement steps

$$(M_{s_1}^{\alpha}, M_{s_1}^{1-\alpha}), \ldots, (M_{s_t}^{1-\alpha}, M_{s_t}^{1-\alpha}),$$

which means that removing the vertices $v_{s_1}, \ldots, v_{s_q}$, the vertex $v_{r_1}$ also will be removed. But with the removal of $v_{l_1}$ the element $e$ returns back to the $\alpha$-domain, which is contradiction.

So, $r > 1$ and the statement (ii) holds. ∇

*Theorem* 6.3. Let a special decomposition be given such that $\{e_{q_1}, \ldots, e_{q_l}\} \subseteq S \setminus M^{\alpha}$.

$R = \{(M_{r_1}^{\alpha}, M_{r_1}^{1-\alpha}), \ldots, (M_{r_r}^{\alpha}, M_{r_q}^{1-\alpha})\}$ is the description of the replaceability procedure associated with the elements $\{e_{q_1}, \ldots, e_{q_l}\}$.

$CG(e_{q_1}, \ldots, e_{q_l})$ is a clean graph associated with the elements $\{e_{q_1}, \ldots, e_{q_l}\}$.

If the set $\{e_{q_1}, \ldots, e_{q_l}\}$ is stable with respect to the cleaning procedure in the graph $G$, then one of the following statements is true:

(i) $\{e_{q_1}, \ldots, e_{q_l}\}$ is an $M^{\alpha}$-reachable set under the given decomposition.

(ii) there are incompatible sets of vertices in the graph $CG(e_{q_1}, \ldots, e_{q_l})$.

*Proof.* Let $v_{c_1}, \ldots, v_{c_k}$ be the vertices of the clean graph $CG(e_{q_1}, \ldots, e_{q_l})$.

The main replacement steps of the procedure corresponding to $R$ are associated with the elements of the set $\{e_{q_1}, \ldots, e_{q_l}\}$, so they will move in the $\alpha$-domain as a result of the replacement procedure. Since the set $\{e_{q_1}, \ldots, e_{q_l}\}$ is stable with respect to the cleaning procedure in the graph $G(e_{q_1}, \ldots, e_{q_l})$, then these elements will also be included in the $\alpha$-domain of the decomposition $(i_1, \ldots, i_k)I(d_nS)$, which correspond to the vertices of the clean graph $CG(e_{q_1}, \ldots, e_{q_l})$.

Let's discus the following cases:

1) all elements included in the $\alpha$-domain of the initial decomposition, are also included in the $\alpha$-domain of the decomposition $(c_1, \ldots, c_k)I(d_nS)$.

Since no element is lost from the $\alpha$-domain of the initial decomposition, and the elements of the set $\{e_{q_1}, \ldots, e_{q_l}\}$ are moved to the (1-$\alpha$)-domain, then in this case:

$$M^{\alpha} \cup \{e_{q_1}, \ldots, e_{q_l}\} \subseteq (c, \ldots, c_k)\, M^{\alpha},$$

which means that the set $\{e_{q_1}, \ldots, e_{q_l}\}$ is reachable under the given decomposition.

2) There are elements that are included in the $\alpha$-domain of the initial decomposition, but are not included in the $\alpha$-domain of the decomposition $(i_1, \ldots, i_k)I(d_n S)$

Recall that due to stability of the set $\{e_{q_1}, \ldots, e_{q_l}\}$ with respect to the cleaning procedure, these elements are included in the $\alpha$-domain of the decomposition $(i_1, \ldots, i_k)I(d_n S)$. This means that some element, $e \notin \{e_{q_1}, \ldots, e_{q_r}\}$, were passed to the (1-$\alpha$)-domain as a result of the replacement or cleaning procedure. That is, there are subsets in the $\alpha$-domain of the initial decomposition, let them be $\{M^{\alpha}_{i_1}, \ldots, M^{\alpha}_{i_k}\}$ and an element $e \in S$ such that the following conditions are satisfied:

- $\forall i \in \{i_1, \ldots, i_k\}\ (e \in M^{\alpha}_{i})$,
- $e \notin (c_1, \ldots, c_p)\, M^{\alpha}$.

By similar reasoning as in the Lemma 6.2, we prove that $k > 1$ and the set

$$\{(M^{\alpha}_{i_1}, M^{1-\alpha}_{i_1}), \ldots, (M^{\alpha}_{i_k}, M^{\alpha}_{i_k})\}$$

is incompatible with respect to the element $e$.

*Corollary* 6.3.1 If the set $\{e_{q_1}, \ldots, e_{q_l}\} = S \backslash M^{\alpha}$ is stable with respect to the cleaning procedure of the graph $G(e_{q_1}, \ldots, e_{q_l})$, the graph $CG(e_{q_1}, \ldots, e_{q_l})$ does not contain incompatible sets of vertices, then for any main replaceability step, $(M^{\alpha}_{i}, M^{1-\alpha}_{i})$, the subset $M^{\alpha}_{i}$ is $M^{\alpha}$-replaceable.

*Proof.* The proof follows from Theorem 6.3. ∇

*Corollary* 6.3.2. If the set $\{e_{q_1}, \ldots, e_{q_l}\} = S \backslash M^{\alpha}$ is stable with respect to the cleaning procedure of the graph $G(e_{q_1}, \ldots, e_{q_l})$, and the graph $CG(e_{q_1}, \ldots, e_{q_l})$ does not contain incompatible sets of vertices, then the $\alpha$-domain of the decomposition corresponding to the graph $CG(e_{q_1}, \ldots, e_{q_l})$ is a special covering for the set $S$.

*Proof.* The proof follows from Theorem 6.3. ∇

Based on Theorem 6.3, it is easy to find incompatible sets of vertices in a graph.

Suppose that as a result of the cleaning procedure in the graph $G(e_{q_1}, \ldots, e_{q_l})$, the stability of the set $\{e_{q_1}, \ldots, e_{q_l}\}$ is not violated. Considering the $\alpha$-domain of decomposition corresponding to the vertices of the clean graph $CG(e_{q_1}, \ldots, e_{q_l})$, we look for elements that were lost from the $\alpha$-domain during the replacement procedure or the cleaning procedure. Then, for each of these elements we find all subsets in the $\alpha$-domain of the initial decomposition containing this element.

If $e$ is such an element and $M_{i_1}^{\alpha}$, . . ., $M_{i_r}^{\alpha}$ are all subsets in the $\alpha$-domain of the decomposition $d_nS$ that contain the element $e$, then $\{v_{i_1}, \ldots, v_{i_r}\}$ is an incompatible set of vertices in the graph $CG(e_{q_1}, \ldots, e_{q_l})$ with respect to the element $e$.

It is easy to notice that:

There cannot be an incompatible set with respect to the active element.

Passive elements do not affect the result of the removal procedure.

6.4. *Extension of the Clean Graph*

Let a special decomposition $d_nS$ be given such that $\{e_{q_1}, \ldots, e_{q_l}\} \subseteq S \setminus M^{\alpha}$.

$G(e_{q_1}, \ldots, e_{q_l})$ is the $M^{\alpha}$-pointed graph corresponding to the description of the replaceability procedure associated with the set $\{e_{q_1}, \ldots, e_{q_l}\}$ under the decomposition $d_nS$,

$CG(e_{q_1}, \ldots, e_{q_l})$ is a clean graph obtaining as a result of applying the cleaning procedure to the graph $G(e_{q_1}, \ldots, e_{q_l})$.

$V_{I_1}, \ldots, V_{I_p}$ are incompatible sets of vertices in the graph $CG(e_{q_1}, \ldots, e_{q_l})$ with respect to the elements $e_{s_1}, \ldots, e_{s_p}$, respectively, such that:

- none of the sets $V_{I_1}, \ldots, V_{I_p}$ contains a removable vertex,

- each of the elements $e_{s_1}, \ldots, e_{s_p}$ is also included in the (1-$\alpha$)-components of some ordered pairs not included in the replaceability procedure $R$. Let them be the ordered pairs:

$$\{(M_{r_1}^{\alpha}, M_{r_1}^{1-\alpha}), \ldots, (M_{r_k}^{\alpha}, M_{r_k}^{1-\alpha})\}.$$

No other incompatible sets are included in the graph $CG(e_{q_1}, \ldots, e_{q_l})$.

We will say that the graph denoted as $ExG(e_{s_1}, \ldots, e_{s_p})$, is an extension of the clean graph $CG(e_{q_1}, \ldots, e_{q_l})$ associated with the elements $e_{s_1}, \ldots, e_{s_p}$ if it is constructed on the basis of the graph $CG(e_{q_1}, \ldots, e_{q_l})$ as follows:

- the graph $ExG(e_{s_1}, \ldots, e_{s_p})$ includes the graph $CG(e_{q_1}, \ldots, e_{q_l})$,

- the ordered pairs $(M_{r_1}^{\alpha}, M_{r_1}^{1-\alpha}), \ldots, (M_{r_k}^{\alpha}, M_{r_k}^{1-\alpha})$ are considered as additional main replacement steps, and the vertices $v_{r_1}, \ldots, v_{r_k}$ corresponding to these steps are considered as additional main vertices for the graph $ExG(e_{s_1}, \ldots, e_{s_p})$,

- new edges and vertices for the graph $ExG(e_{s_1}, \ldots, e_{s_p})$ are constructed using the description of the replaceability procedure started from the replacement steps

$$(M_{r_1}^{\alpha}, M_{r_1}^{1-\alpha}), \ldots, (M_{r_k}^{\alpha}, M_{r_k}^{1-\alpha})$$

under the decomposition $d_nS$.

The vertices and edges that appeared during this procedure will not be added to the new graph, if they are already included in it or were removed during any removal procedures.

If no incompatible set satisfies the described conditions, then we assume that the graph $ExG(e_{s_1}, \ldots, e_{s_p})$ coincides with the graph obtained as a result of compatibility procedure, and the set $\{e_{s_1}, \ldots, e_{s_p}\}$ is empty.

*The set* $\{e_{s_1}, \ldots, e_{s_p}\}$ *will also be called the set of graph extension elements.*

The graph $ExG(e_{s_1}, \ldots, e_{s_p})$ will be called the final extension for the clean graph $CG(e_{q_1}, \ldots, e_{q_l})$ if there are no more elements that extend the graph.

In other words, vertices end edges enter the graph $ExG(e_{s_1}, \ldots, e_{s_p})$ if and only if they enter the extending graph or appear in the replaceability procedure started from the main steps

$$(M^{\alpha}_{r_1}, M^{1-\alpha}_{r_1}), \ldots, (M^{\alpha}_{r_k}, M^{1-\alpha}_{r_k})$$

under the decomposition $d_n S$.

6.5. *Incompatibility Elimination Procedure in the Graph.*

The procedure described below will be called the vertex compatibility procedure or the procedure for eliminating the vertex incompatibility in a clean graph. It consists in the sequential application of the removal procedure to the vertices of incompatible sets in the graph as follows:

Let $CG(e_{q_1}, \ldots, e_{q_l})$ be a clean graph, and $V_I$ an incompatible set of vertices in it with respect to some element $e_r \in S$.

(i) we apply the removal procedure to some vertex $v \in V_I$.

If the vertex $v$ is not removable, we consider the initial graph and apply the removal procedure to another vertex of the set $V_I$.

The application of the removal procedure to the vertices included in $V_i$ is suspended if a removable vertex is found in it.

In this case, we say that the incompatibility of the set $V_I$ is eliminated and proceed to consider other incompatible sets, if any exists.

(ii) let $V_I$ be an incompatible set with respect to the element $e_r$ such that none of vertices in it is removable.

In this case we check whether the element $e_r$ is included in the (1-$\alpha$)-components of some ordered pairs not included in the replaceability procedure.

(ii.1) If $e_r$ is not included in any of the (1-$\alpha$)-components of these ordered pairs, we conclude that the incompatibility of the set $V_I$ is not eliminated. This means that the graph vertex incompatibility cannot be eliminated, so we suspend the procedure.

(ii.2) If $e_r$ is included in the (1-$\alpha$)-components of some of these ordered pairs, then

- we mark $e_r$ as an element which will be considered later, and continue considering other vertices if this incompatible set.

- we continue apply the removal procedure to the vertices of the other incompatible set, if any exists.

(iii) Let, as a result of applying the compatibility procedure to all incompatible sets in the graph, we have the following:

- the incompatibilities of some sets are eliminated,

- the incompatibilities of other sets are not eliminated. At the same time, these sets are incompatible with respect to the elements, let them be the elements $e_{s_1}, \ldots, e_{s_p}$, each of which is also included in (1-$\alpha$)-components of some ordered pairs not included in the replaceability procedure.

In this case we construct the graph $ExG(e_{s_1}, \ldots, e_{s_p})$ as an extension of the clean graph under consideration, using Description 6.4.

In order to study the $M^{\alpha}$-reachability of the set $\{e_{s_1}, \ldots, e_{s_p}\}$ under the decomposition corresponding to the vertices of the graph $ExG(e_{s_1}, \ldots, e_{s_p})$, we apply cleaning procedure to the graph $ExG(e_{s_1}, \ldots, e_{s_p})$ if the $\alpha$-domain of this decomposition does not cover the set $S$.

If the set $\{e_{q_1}, \ldots, e_{q_l}, e_{s_1}, \ldots, e_{s_p}\}$ is stable with respect to the cleaning procedure in the graph, but the $\alpha$-domain of the decomposition corresponding to the vertices of the resulting graph does not cover the set $S$, then

- we find the incompatible sets of vertices in the resulting graph,

- we apply the incompatibility elimination procedure to these incompatible sets.

We continue the extension of the graph, exploring new incompatible sets and adding new main vertices, if they appear during these procedures, according to this description.

*We will say that the incompatibility of the vertices in the graph* $CG(e_{q_1}, \ldots, e_{q_l})$ *is eliminated if* $e_{q_1}, \ldots, e_{q_l}$ *and all elements due to which the graph was expanded, are stable with respect to all cleaning and removal procedures, and new incompatible sets are not appear.*

Let $V_{I_1}, \ldots, V_{I_k}$ be incompatible sets of vertices in the clean graph $CG(e_{q_1}, \ldots, e_{q_l})$.

We will use the notation $(V_{I_1}, \ldots, V_{I_k})CG$ for the graph defined as follows:

- if the graph in question is not extended when applying the compatibility procedure to it, then $(V_{I_1}, \ldots, V_{I_k})CG$ is the graph obtained as a result of applying the compatibility procedure to the sets $V_{I_1}, \ldots, V_{I_k}$ in any order.

- if the graph in question has been extended when applying the compatibility procedure to it, then the graph $ExG$ will be considered as the graph $(V_{I_1}, \ldots, V_{I_k})CG$.

If any of these incompatible sets is not eliminated, then we consider $(V_{I_1}, \ldots, V_{I_k})CG$ as an empty graph.

For any main vertex $v_i$, the subgraph of the graph $(V_{I_1}, \ldots, V_{I_k})CG$ corresponding to the graph $CG[v_i]$ will be denoted by $(V_{I_1}, \ldots, V_{I_k})CG[v_i]$.

If the main vertex $v_i$ is removed during the compatibility procedure, then we consider $(V_{I_1}, \ldots, V_{I_k})CG[v_i]$ as an empty subgraph.

*Proposition* 6.6. If $CG(e_{q_1},\dots,e_{q_l})$ is a clean graph associated with the set $\{e_{q_1},\dots,e_{q_l}\}$, and $V_I$ is an incompatible set of vertices containing a removable vertex, then:

For any main vertex, $v \in (V_I)CG$, the subgraph $(V_I)CG[v]$ is a connected graph, and there are paths from the vertex $v$ to any of the other vertices.

*Proof.* According to Lemma 4, for any $v \in CG[v]$ is a connected graph, and there are paths from the vertex $v$ to any of the other vertices. The compatibility procedure actually consists of the removal procedures, and these procedures preserve the connectivity of the graph. Since $v \in (V_I)CG$, then $v$ is not removed during the removal procedure. So, $(V_I)CG[v]$ is a connected graph. ∇

*Definition* 6.7. Let a special decomposition is given such that

$$\{e_{q_1}, \dots, e_{q_l}\} \subseteq S\backslash M^{\alpha},$$

and the set of elements $\{e_{q_1}, \dots, e_{q_l}\}$ is a stable with respect to the cleaning procedure in the graph $G(e_{q_1}, \dots, e_{q_l})$.

We say, that the set $\{e_{q_1}, \dots, e_{q_l}\}$ is stable with respect to some incompatible set of vertices, if the incompatibility associated with this set can be eliminated.

We say that the set $\{e_{q_1}, \dots, e_{q_l}\}$ is stable with respect to the compatibility procedure in the graph, if the incompatibility of any set of vertices can be eliminated.

*Theorem* 6.8. Let $CG(e_{q_1}, \dots, e_{q_l})$ be a clean graph associated with the set $\{e_{q_1}, \dots, e_{q_l}\}$, and $ExG(e_{s_1}, \dots, e_{s_p})$ is its extension associated with the elements $e_{s_1}, \dots, e_{s_p}$.

If there is an element

$$e \in \{e_{q_1},\dots,e_{q_l}, e_{s_1}, \dots, e_{s_p}\}$$

that is not stable when the cleaning or compatibility procedure is applied to the graph

$$ExG(e_{s_1}, \dots, e_{s_p}),$$

then there is no $M^{\alpha}$-covering for the set $S$ under the initial decomposition.

*Proof.* Since $ExG(e_{s_1}, \dots, e_{s_p})$ is extension of the graph $CG(e_{q_1}, \dots, e_{q_l})$, then there are sets of vertices in the graph, denote them as $V_{I_1}, \dots, V_{I_p}$, such that:

- $V_{I_1}, \dots, V_{I_p}$ are incompatible with respect to the elements $e_{s_1}, \dots, e_{s_p}$, respectively,

- none of these sets contains a removable vertex,

- each of the elements $e_{s_1}, \dots, e_{s_p}$ is also included in the (1-$\alpha$)-components of some ordered pairs not included in the replaceability procedure.

Let them be the pairs:

$$\{(M^{\alpha}_{r_1}, M^{1-\alpha}_{r_1}), \dots, (M^{\alpha}_{r_k}, M^{1-\alpha}_{r_k})\}.$$

In fact, we are studying the possibility of moving the element $e$ to the $\alpha$-domain along with the elements $\{e_{q_1}, \dots, e_{q_l}\}$. This cannot be done using only the vertices of the clean graph $CG(e_{q_1}, \dots, e_{q_l})$, since the sets $V_{I_1}, \dots, V_{I_p}$ consist only of unremovable vertices. Therefore, we construct the graph $ExG(e_{s_1}, \ldots, e_{s_p})$ using the graph $CG(e_{q_1}, \ldots, e_{q_l})$ and the replaceability procedure started from the steps

$$\{(M_{r_1}^{\alpha}, M_{r_1}^{1-\alpha}), \ldots, (M_{r_k}^{\alpha}, M_{r_k}^{1-\alpha})\}.$$

Constructing the graph $ExG(e_{s_1}, \ldots, e_{s_p})$, the previously removed vertices of the graph $CG(e_{q_1}, \ldots, e_{q_l})$ will not be restored, since in any case they would have been removed from the new graph during the same removal procedure.

The main vertices of the graph $ExG(e_{s_1}, \ldots, e_{s_p})$ are associated with the elements

$$\{e_{q_1}, \ldots, e_{q_l}, e_{s_1}, \ldots, e_{s_p}\}.$$

At the same time no other vertices are associated with these elements.

Thus, it is easy to see that if $e$ is an element of this set,

$$e \in \{e_{q_1}, \ldots, e_{q_l}, e_{s_1}, \ldots, e_{s_p}\},$$

then $e$ will move to the $\alpha$-domain only if any of the main vertices associated with $e$ is not removed. So, if all the main vertices associated with $e$ are removed during the removal procedures, the initial decomposition does not contain a special covering for the set $S$. ∇

*Theorem* 6.9. If $CG(e_{q_1}, \ldots, e_{q_l})$ is a clean graph associated with the set $\{e_{q_1}, \ldots, e_{q_l}\}$,

$V_I$ is an incompatible set of vertices in the graph with respect to some element $e$,

The set $\{e_{q_1}, \ldots, e_{q_l}\}$ is stable as a result of applying the compatibility procedure to the set $V_I$, and $\{v_{i_1}, \ldots, v_{i_p}\}$ is the set of vertices of the graph $(V_I)CG$,

Then, the set $(i_1, \ldots, i_p)M^{\alpha}$ contains all active elements of the replaceability procedure associated with the set $\{e_{q_1}, \ldots, e_{q_l}\}$.

*Proof.* Suppose that the set $V_I$ contains a removable vertex. Let it be the vertex $v$. Applying the removal procedure to the vertex $v$, we get the graph $(V_I)CG$. In this case the compatibility procedure is actually a removal procedure. So, according to the Theorem 5.4, the set $(i_1, \ldots, i_p)M^{\alpha}$ contains all active elements of the replaceability procedure associated with the set $\{e_{q_1}, \ldots, e_{q_l}\}$.

Let all vertices included in the set $V_I$ are not removable, and also, the element $e$ is included in (1-$\alpha$)-components of some ordered pairs not included in the replaceability procedure associated with the set $\{e_{q_1}, \ldots, e_{q_l}\}$.

In this case, we construct the graph $ExG(e)$ and apply the cleaning procedure to it if useless vertices appeared in it.

The construction procedure of $ExG(e)$ consists of replaceability procedure, that is not related to the existing replacement steps. That is, the replacement steps corresponding to the vertices of the clean graph $CG$, are not changed. So, at this stage the active elements participated in the replaceability procedure associated with the set $\{e_{q_1}, \ldots, e_{q_l}\}$ are not lost from the $\alpha$-domain.

If useless vertices appeared in the graph $ExG(e)$, we apply the cleaning procedure to it. Since the set $\{e_{q_1}, \ldots, e_{q_l}\}$ is a stable with respect to the compatibility procedure for the set $V_I$, then obviously it will be stable also with respect the cleaning procedure. So, according to the Corollary 5.4.1, the set $(i_1, \ldots, i_p)M^{\alpha}$ contains all active elements participated in the replaceability procedure associated with the set $\{e_{q_1}, \ldots, e_{q_l}\}$. ∇

*Theorem* 6.10. Let $\{e_{q_1}, \ldots, e_{q_l}\}$ be a stable set with respect to the cleaning procedure in the graph $G(e_{q_1}, \ldots, e_{q_l})$, and $CG(e_{q_1}, \ldots, e_{q_l})$ a corresponding clean graph.

The total number of incompatible sets that appeared as a result of the replacement and the removal procedures, when applying the incompatibility elimination procedure to the clean graph, is less than the number of elements of the set $S$.

*Proof*. Suppose that $V_{I_1}, \ldots, V_{I_p}$ are incompatible sets of vertices in a clean graph with respect to the elements $e_{s_1}, \ldots, e_{s_p}$, respectively. Obviously, the number of these sets, $p$, is less than the number of elements of the set $S$.

We prove that if, as a result of successive application of the removal procedure, incompatible sets appear, then the elements with respect to which these sets are incompatible do not repeat.

According to the incompatibility elimination procedure, we successively apply the removal procedure to the vertices of the sets $V_{I_1}, \ldots, V_{I_p}$.

(i) Let none of the elements $e_{s_1}, \ldots, e_{s_p}$ be included in (1-$\alpha$)-components of the ordered pairs not included in the replaceability procedure.

In this case, if there is an incompatible set $V_I \in \{V_{I_1}, \ldots, V_{I_p}\}$ that do not contain a removable vertex, then the incompatibility elimination procedure in the graph is suspended, which means that no new incompatible set appears.

Suppose that each of these incompatible sets $V_{I_1}, \ldots, V_{I_p}$ contain a removable vertex, let them be the vertices $v_{i_1}, \ldots, v_{i_p}$, respectively. Applying the removal procedure to each of these vertices the incompatibility of the corresponding set will be eliminated. In the same time, the vertices $v_{i_1}, \ldots, v_{i_p}$ will be removed, which means that the elements $e_{s_1}, \ldots, e_{s_p}$ will move to the $\alpha$-domain of the corresponding decomposition.

If, as a result of these removal procedures new incompatible sets of vertices appear, then none of the vertices $v_{i_1}, \ldots, v_{i_p}$ will be included in any of them since these vertices are removed. Therefore, new sets will be incompatible with respect to the elements that are differ than th elements $e_{s_1}, \ldots, e_{s_p}$.

(ii) Let's consider the case when some incompatible sets contain removable vertices, while in the rest of the sets none of the vertices is removable.

Assume that $V_{j_1}, \ldots, V_{j_k}$ are incompatible sets with respect to the elements $e_{l_1}, \ldots, e_{l_k}$, respectively, such that none of them contains removable vertices. In the same time, the elements $e_{l_1}, \ldots, e_{l_k}$ are also included in (1-$\alpha$)-components of some ordered pairs not included in the replaceability procedure.

In this case, after eliminating the incompatibilities of the sets containing removable vertices, the graph $ExG(e_{l_1}, \ldots, e_{l_k})$ is constructed as an extension of the resulting graph, if the sets $V_{j_1}, \ldots, V_{j_k}$ are still incompatible.

Suppose that as a result of replacement procedure, new incompatible sets appear with respect to some elements. According to description 6.4, the vertices removed during the procedures for eliminating previous incompatible sets do not enter the graph $ExG(e_{l_1}, \ldots, e_{l_k})$. Therefore, these vertices cannot be included in any new incompatible set, which means that the new sets are incompatible with respect to new elements.

As for the elements $e_{l_1}, \ldots, e_{l_k}$, for each of them there are main vertices in the graph $ExG(e_{l_1}, \ldots, e_{l_k})$ associated with it. That is, these elements are restored in the $\alpha$-domain by means of corresponding main vertices. Therefore, if as a result of some removal procedure any of these elements is unstable, the procedure is suspended, since there cannot be a special covering for the set $S$.

This means that no extension of this graph can contain a new set of vertices that is incompatible with respect to one of these elements.

Therefore, if the incompatibility of the set with respect to some element has been eliminated, then another incompatible set with respect to the same element cannot appear as a result of the removal or replacement procedures.

Obviously, a set can be incompatible with respect to several elements. But different sets of vertices in the same graph cannot be incompatible with respect to the same element.

Thus, the total number of incompatible sets that appeared as a result of the replacement and the removal procedures in the graph, is less than the number of elements of the set $S$. ∇

The theorems 6.8, 6.9 and 6.10 are important in they give a clear direction for finding out the possibility of eliminating the incompatibility of graph vertices.

That is, by Theorem 6.8, if the graph is extended, the elements of extension create new main vertices, and thus, they acquire the significance of the elements that created the main vertices in the initial graph.

We apply the compatibility procedure for restoring the lost elements. Unfortunately, new incompatible vertices may appear during this procedure. By Theorem 6.9, new elements that create new incompatible sets were not active in the previous decomposition. Only passive elements create new incompatible sets and thus become active in the new decomposition.

The Theorem 6.10 estimates the upper bound on the number of incompatible sets that arise in the graph during all procedures.

*Lemma* 6.11 Let the special decomposition $d_nS$ be given such that

$$\{e_{q_1}, \ldots, e_{q_l}\} \subseteq S \backslash M^\alpha,$$

$\{e_{q_1}, \ldots, e_{q_l}\}$ is a stable set of elements with respect to the cleaning procedure in the graph $G(e_{q_1}, \ldots, e_{q_l})$.

$V_1$ and $V_2$ are incompatible sets of vertices in the clean graph $CG(e_{q_1}, \ldots, e_{q_l})$.

(i) If the incompatibility of the set $V_1$ cannot be eliminated in the graph $CG$, and the incompatibility of the set $V_2$ can be eliminated in the graph $CG$, then the incompatibility of the set $V_1$ also cannot be eliminated in the graph $(V_2)CG$.

(ii) If the incompatibility each of the sets $V_1$ and $V_2$ can be eliminated in the graph $CG$, then the incompatibility of the set $V_2$ is eliminated in the graph $(V_1)CG$ if and only if the incompatibility of the set $V_1$ is eliminated in the graph $(V_2)CG$.

*Proof.* (i) Suppose that it is impossible to eliminate the incompatibility of the set $V_1$ in the graph $CG$, and the incompatibility of the set $V_2$ can be eliminated.

Since the incompatibility of the set $V_1$ cannot be eliminated in the graph $CG$, then as a result of applying the removal procedure to any vertex of the set $V_1$, the stability of the elements $\{e_{q_1}, \ldots, e_{q_l}\}$ is violated.

Let $P(v)$ denote the set of all vertices that are removed as a result of applying the removal procedure to the vertex $v$.

Suppose that the incompatibility of the set $V_2$ is eliminated by applying the removal procedure to some vertex $v \in V_2$. Obviously, none of the vertices of the set $V_1$ is included in the set $P(v)$ as otherwise the continuing procedure after removing an element of the set $V_1$ will break the stability of the set $\{e_{q_1}, \ldots, e_{q_l}\}$, and the incompatibility of $V_2$ will not be eliminated.

Consider the following cases for an arbitrary vertex $u \in V_1$:

(1) $P(v) \cap P(u) = \emptyset$.

This means that all vertices of the set $P(u)$ are included in the graph $(V_2)CG$. As for the edges that are removed as a result of application of the removal procedure to the vertex $u$, some of them may be removed in the removal procedure applied the vertex $v$. Obviously, the remaining edges will be removed as a result of applying the removal procedure to the vertex $u$ in the graph $(V_2)CG$. Therefore, this procedure will not eliminate the incompatibility of the set $W_1$ in the graph $(V_2)CG$.

(2) $P(v) \cap P(u) \neq \emptyset$.

That is, there are common vertices that are removed both during the removal procedure launched from the vertex $v$ in the graph $CG$, and during the removal procedure launched from the vertex $u$ in the graph $CG$.

Recall that if some vertex is removed both as a result of applying the removal procedure to the vertex $v$ and as a result of applying the removal procedure to the vertex $u$, then the actions after removing this vertex are the same in both procedures.

By assumption the incompatibility of the set $V_2$ is eliminated as a result of applying the removal procedure to the vertex $v$. So, the removal only of common vertices from the set $P(u)$ do not violate the stability of the set $\{e_{q_1}, \ldots, e_{q_l}\}$. This means that the stability of this set will be violated if other vertices from the set $P(u)$ are also removed.

Consider the removal procedure in the graph $(V_2)CG$ started from the vertex $u$. Recall that $u \notin P(v)$ and the common vertices included also in the set $P(u)$ have already been removed. In the course of this procedure, all the remaining vertices of the set $P(u)$ will be removed. So, all vertices of the set $P(u)$ are removed, which violates the stability of the set $\{e_{q_1}, \ldots, e_{q_l}\}$.

(ii) Suppose that the incompatibility each of the sets $V_1$ and $V_2$ can be eliminated in the graph $CG$, as well as the incompatibility of the set $V_2$ is eliminated in the graph $(V_1)CG$.

Consider arbitrary vertices $u \in V_1$ and $v \in V_2$ for which the following is true:

- the incompatibility of the set $V_1$ in the graph $CG$ is eliminated as a result of applying the removal procedure to the vertex $u$.
- the incompatibility of the set $V_2$ in the graph $CG$ is eliminated as a result of applying the removal procedure to the vertex $v$.

Suppose that none of the vertices of the set $V_1$ has been removed as a result of applying the removal procedure to the vertex $v$. That is, as a result of compatibility procedure for the set $V_2$, the set $V_1$ is still incompatible.

Let us apply the removal procedure to the vertex $u$ in the graph $(V_2)CG$ and assume that this procedure violates the stability of the set $\{e_{q_1}, \ldots, e_{q_l}\}$.

Since both removal procedures in the graph $CG$ launched from the vertex $u$ and from the vertex $v$ separately do not violate the stability of the set $\{e_{q_1}, \ldots, e_{q_l}\}$, then there are vertices $v_{p_1}, \ldots, v_{p_r}$ such that:

- $v_{p_1}, \ldots, v_{p_r}$ are not removed during both separate procedures,
- $v_{p_1}, \ldots, v_{p_r}$ are removed as a result of applying the removal procedure to the vertex $u$ in the graph $(V_1)CG$.
- removal of these vertices leads to violation of the stability of the set $\{e_{q_1}, \ldots, e_{q_l}\}$.

It is easy to see that this can only take place if each of these vertices has outgoing disjunctive edges associated with the same element, so that some of the outgoing disjunctive edges are removed during the removal procedure launched from the vertex $v$ in the graph $G$, and others - during the removal procedure launched from the vertex $u$ in the graph $(V_1)CG$.

It is obvious that if these procedures are performed separately, none of these vertices will be removed. It is also obvious that as a result of successive procedures launched from the vertex $u$ and launched from the vertex $v$, these vertices will be removed.

But obviously, these vertices will be removed also as a result of successive procedures launched from the vertex $v$ and launched from the vertex $u$, which means that the procedure launched from the vertex $v$ does not eliminate in the incompatibility of the set $V_2$ in the graph $(V_1)CG$. We get a contradiction. So, the point (ii) is true. ∇

*Corollary* 6.11.1. Let the special decomposition $d_nS$ be given such that

$$\{e_{q_1}, \ldots, e_{q_l}\} \subseteq S \backslash M^{\alpha},$$

$\{e_{q_1}, \ldots, e_{q_l}\}$ is a stable set of elements with respect to the cleaning procedure in the graph $G(e_{q_1}, \ldots, e_{q_l})$.

$V_1$ and $V_2$ are incompatible sets of vertices in the clean graph $CG(e_{q_1}, \ldots, e_{q_l})$.

(i) If the incompatibility of the set $V_1$ can be eliminated in the graph $CG$, and the incompatibility of the set $V_2$ cannot be eliminated in the graph $(V_1)CG$, then the incompatibility of the set $V_2$ also cannot be eliminated in the graph $CG$.

*Proof.* Suppose that under the given conditions the incompatibility of the set $V_2$ is eliminated in the graph $CG$. Obviously, the conditions of Lemma 6.11 (i) are satisfied. This means that $V_2$ will be eliminated in the graph $(V_1)CG$, which is contradiction. ∇

*Theorem* 6.12. Let $d_nS$ be a special decomposition of the set $S$ such that

$$\{e_{q_1}, \ldots, e_{q_l}\} \subseteq S \backslash M^{\alpha}.$$

$\{e_{q_1}, \ldots, e_{q_l}\}$ is a stable set with respect to the cleaning procedure in the pointing graph $G(e_{q_1}, \ldots, e_{q_l})$.

If there are different incompatible sets of vertices in the graph $CG(e_{q_1}, \ldots, e_{q_l})$ then the result of the compatibility procedures associated with these sets does not depend on the sequence in which they are considered.

*Proof.* Suppose that $V_{i_1}, \ldots, V_{i_k}$ are incompatible sets in the graph $CG(e_{q_1}, \ldots, e_{q_l})$ and $V_{i_j}$ is any of these sets $V_{i_j} \in \{V_{i_1}, \ldots, V_{i_r}\}$.

If the incompatibility of the set $V_{i_j}$ cannot be eliminated in the graph $CG(e_{q_1}, \ldots, e_{q_l})$, then by Lemma 6.11 (i), the incompatibility $V_{i_j}$ cannot be eliminated also in the graph $(V_{i_k})CG$ for any incompatible set $V_{i_k} \in \{V_{i_1}, \ldots, V_{i_r}\}$ such that the incompatibility of $V_{i_k}$ is eliminated in the graph $CG(e_{q_1}, \ldots, e_{q_l})$. So, in this case, the incompatibility in the graph cannot be eliminated by considering these sets in any order.

Let the following conditions hold:

- the incompatibility of each of the sets $\{V_{j_1}, \ldots, V_{j_p}\} \subseteq \{V_{i_1}, \ldots, V_{i_r}\}$ is eliminated in the graph $CG(e_{q_1}, \ldots, e_{q_l})$,

- $(V_{i_1}, \ldots, V_{i_p})CG$ is not an empty graph,

- the incompatibility of the set $V_{i_j} \in \{V_{i_1}, \ldots, V_{i_r}\} \backslash \{V_{j_1}, \ldots, V_{j_p}\}$, is not eliminated in the graph $(V_{i_1}, \ldots, V_{i_r})CG$.

According to Lemma 6.11, and Corollary 6.11.1, this means that the incompatibility of the set $V_{i_q}$ also cannot be eliminated in the graph $(V_{i_1}, \ldots, V_{i_{r-1}})CG$. By the same reasoning, we obtain that the incompatibility of the set $V_{i_q}$ cannot be eliminated in $CG(e_{q_1}, \ldots, e_{q_l})$.

Thus, using also the Lemma 6.11 (ii), we can conclude that the result of the compatibility procedures for the sets $V_{i_1}, \ldots, V_{i_r}$ does not depend on the sequence in which these sets are considered. ∇

*Theorem* 6.13. Let a special decomposition $d_nS$ of the set $S$ be given such that

$$\{e_{q_1}, \ldots, e_{q_l}\} = S \backslash M^{\alpha},$$

$G(e_{q_1}, \ldots, e_{q_l})$ is the pointing graph associated with the set $\{e_{q_1}, \ldots, e_{q_l}\}$.

$ExG(e_{s_1}, \ldots, e_{s_p})$ is the final extension of the clean graph $CG(e_{q_1}, \ldots, e_{q_l})$.

There is a special covering for the set $S$ in the decomposition $d_nS$ if and only if the set

$$\{e_{q_1}, \ldots, e_{q_l}, e_{s_1}, \ldots, e_{s_p}\}$$

is stable with respect to both the cleaning procedure and the compatibility procedure in the graph $ExG(e_{s_1}, \ldots, e_{s_p})$.

*Proof.* Recall that if the graph $CG(e_{q_1}, \ldots, e_{q_l})$ does not extended, then the graph $ExG(e_{s_1}, \ldots, e_{s_p})$ coincides with the clean graph obtained as a result of compatibility procedure, and the set $\{e_{s_1}, \ldots, e_{s_p}\}$ is empty.

Suppose that during the search for $M^{\alpha}$-reachability of the set $\{e_{q_1}, \ldots, e_{q_l}\}$, the corresponding clean graph is extended due to the elements $e_{s_1}, \ldots, e_{s_p}$, which form new main replacement steps. In the course of these procedures, we obtain:

- new special decomposition, since the elements $e_{s_1}, \ldots, e_{s_p}$ may appear during various compatibility procedures,

- new description of the replaceability procedure, as we have expanded the set of main replaceability steps.

Since $ExG(e_{s_1}, \ldots, e_{s_p})$ is the final extension of the clean graph, we will search for reachability of the set

$$\{e_{q_1}, \ldots, e_{q_l}, e_{s_1}, \ldots, e_{s_p}\}$$

under the new decomposition.

Obviously, the mentioned new decomposition is an $I$-transformation of $d_nS$, so any special covering of the set $S$ under this decomposition will also be a special covering under the decomposition $d_nS$ and vice versa.

By Lemma 1.8 in [13], every special covering for the set $S$ is also an $M^{\alpha}$-covering for the set $S$.

Let for some $\alpha_1, \alpha_2, \ldots, \alpha_n$, $(\alpha_i \in \{0,1\})$, the set

$$c_nS = \{M_1^{\alpha_1}, \ldots, M_i^{\alpha_i}, \ldots, M_n^{\alpha_n}\}$$

be a special covering for the set $S$ and, in addition, $c_nS$ coincides with the $\alpha$-domain of the corresponding decomposition, that is, $c_nS$ is an $M^\alpha$-covering for the set $S$.

By Theorem 3.5 and Corollary 3.5.1, this means that the set

$$\{e_{q_1}, \ldots, e_{q_l}, e_{s_1}, \ldots, e_{s_p}\}$$

is a reachable. Since as a result of all cleaning and compatibility procedures, these elements are found in the $\alpha$-domain of the corresponding decomposition, then the set

$$\{e_{q_1}, \ldots, e_{q_l}, e_{s_1}, \ldots, e_{s_p}\}$$

is stable with respect of all these procedures.

The other side:

Suppose that the set

$$\{e_{q_1}, \ldots, e_{q_l}, e_{s_1}, \ldots, e_{s_p}\}$$

is stable with respect to both the cleaning procedure and the compatibility procedure in the graph $ExG(e_{s_1}, \ldots, e_{s_p})$.

Recall that $ExG(e_{s_1}, \ldots, e_{s_p})$ is the final extension of the graph.

This means that:

- any incompatible set in it contains a removable vertex,
- if new incompatible sets arise as a result of performing procedures, then they will be eliminated,
- as a result of all required procedures, all elements will be found in the $\alpha$-domain of the resulting decomposition.

So, the $\alpha$-domain of this decomposition will be a special covering for the set $S$. ∇

The search for a special covering for a set under a given decomposition consists of various algorithmic procedures. The Theorem 6.13 describes a deterministic sequence of required procedures that regulate such a search.

In fact, based on a special decomposition, we build a graph and strive to eliminate useless vertices and incompatible sets of vertices in it.

In some cases, during the procedure, we extend the graph and continue the same actions in the new graph.

As a result of all these procedures, either we obtain a special covering for the set or we conclude that there is no special covering for the considered set under given decomposition.

6.14. *Finding the maximum number of elements included in the α-domain.*

An important question arises if, under the given decomposition, there does not exist a special covering for the set $S$:

*What is the maximum number of elements that can simultaneously be in the* $\alpha$*-domain of the resulting decomposition and how to assert that the* $\alpha$*-domain of the obtained decomposition contains the maximum number of elements of the set* $S$.

Assume that as a result of all cleaning and compatibility procedure in the corresponding graphs, we obtain the decomposition $(i_1,\ldots,i_k)I(d_nS)$ in which there are elements, let them be $e_{r_1},\ldots, e_{r_s}$, that are included only in the (1- $\alpha$)-domain of this decomposition. That is,

$$\{e_{r_1},\ldots, e_{r_s}\} = S \setminus (i_1,\ldots, i_k)\, M^{\alpha}.$$

This means that:

- for any $e_r \in \{e_{r_1},\ldots, e_{r_s}\}$, there is an incompatible set of vertices with respect to $e_r$.
- none of these incompatible sets contains a removable vertex.

So, only the elements of the set $S \setminus \{e_{r_1},\ldots, e_{r_s}\}$ are included in the $\alpha$-domain.

It is obvious, that the replaceability procedure cannot solve this problem. This means that we need to describe another procedure, as a result of which we get needed decomposition.

Suppose that the elements $\{e_{r_1},\ldots, e_{r_s}\}$ are distributed over the subsets

$$M_{i_1}^{1-\alpha},\ldots, M_{i_r}^{1-\alpha}$$

in the (1-$\alpha$)-domain of the corresponding decomposition.

It is easy to see, that if $M_i^{\alpha} \in \{M_{i_1}^{\alpha},\ldots, M_{i_r}^{\alpha}\}$, then $M_i^{\alpha}$ is an $M^{\alpha}$-single subset with respect to some elements. Otherwise, by means of permutation of components of the ordered pair ($M_i^{\alpha}$, $M_i^{1-\alpha}$), the corresponding element of the set $\{e_{r_1},\ldots, e_{r_s}\}$ will move to the $\alpha$-domain, and no element is lost from the $\alpha$-domain. But then the incompatibility of the corresponding set will be eliminated, which will contradict the assumption.

Consider the set of ordered pairs

$$\{(M_{i_1}^{\alpha}, M_{i_1}^{1-\alpha}),\ldots, (M_{i_r}^{\alpha}, M_{i_r}^{1-\alpha})\}.$$

Obviously, ordered pairs not included in this set cannot contribute to solving the problem, since as a result of a permutation of the components of any of them, the $\alpha$-domain will not receive a new element, but may lose an element. Therefore, all actions should be performed only within this set.

As for ordered pair included in this set, that is, if

$$(M_i^{\alpha}, M_i^{1-\alpha}) \in \{(M_{i_1}^{\alpha}, M_{i_1}^{1-\alpha}),\ldots, (M_{i_r}^{\alpha}, M_{i_r}^{1-\alpha})\},$$

then we permute the components of the pair ($M_i^{\alpha}$, $M_i^{1-\alpha}$) only if the number of elements in $M_i^{\alpha}$ with respect to which it is $M^{\alpha}$-single, is less than the number of elements of the set

$$M_i^{1-\alpha} \cap \{e_{r_1},\ldots, e_{r_s}\}.$$

Obviously, these permutations do not constitute a replaceability procedure.

During the permutation of the components of any of these ordered pairs, the $\alpha$-domain loses an element, which is not restored. At the same time, $\alpha$-domain receives more new elements than it loses.

So, the resulting decomposition will contain a maximum number of elements.

## 7. Boolean Functions in Conjunctive Normal Form and Special Decompositions

Let for natural numbers $n$ and $m$, $f(x_1, x_2, \ldots, x_n)$ be a Boolean function of $n$ variables represented in conjunctive normal form ($CNF$) with $m$ clauses.

We denote by $c_i$ the $i$-th clause of the formula in a certain natural numbering of the clauses. That is for some $k \in \{1, \ldots, n\}$ and $\{ j_1, \ldots, j_k\} \subseteq \{1, \ldots, n\}$, we denote

$$c_i = x_{j_1}^{\alpha_1} \vee \ldots \vee x_{j_k}^{\alpha_k} ,$$

where $\alpha_j \in \{0,1\}$, $x_j^0 = \neg x_j$, $x_j^1 = x_j$, $j \in \{1, \ldots, n\}$.

With this notation, the function $f(x_1, x_2, \ldots, x_n)$ will be represented as

$$f(x_1, x_2, \ldots, x_n) = \wedge_{i=1}^{m} c_i.$$

For simplicity and technical convenience, we assume that

- no variable and its negation are included in any clause simultaneously,

- if the function contains $n$ variables, then they are numbered sequentially. That is, for every $j \in \{1, \ldots, n\}$, the literal $x_j^{\alpha}$ appears in some clauses for some $\alpha \in \{0,1\}$.

Obviously, this assumption does not limit the set of functions being considered.

*We say that the set of clauses* $\{c_{j_1}, \ldots, c_{j_k}\}$ *is satisfiable if there is a tuple* $(\sigma_1, \ldots, \sigma_n)$, *where* $\sigma_j \in \{0,1\}$ *such that any of these clauses takes 1 when giving the values* $\sigma_1, \ldots, \sigma_n$ to *the variables* $x_1, \ldots, x_n$, *respectively.*

The $satCNF$ problem is the problem of determining if given function $f$ represented in $CNF$ is satisfiable, that is if there exist $\sigma_1, \ldots, \sigma_n$ such that $\sigma_j \in \{0,1\}$ and

$$f(\sigma_1, \sigma_2 \ldots, \sigma_n) = 1.$$

We will use the following notations:

$S(f)$ is the set of clauses of the function $f(x_1, \ldots, x_n)$:

$$S(f) = \{c_1, c_2, \ldots, c_m\}.$$

For each $i \in \{1, \ldots, n\}$ and $\alpha \in \{0,1\}$ we compose the sets $fM_i^{\alpha}$ and $fM_i^{1-\alpha}$ as follows:

$$fM_i^{\alpha} = \{c_j \,/\, c_j \in S(f) \text{ and } c_j \text{ contains the literal } x_i^{\alpha}, \ (j \in \{1, \ldots, m\})\}.$$

$$fM_i^{1-\alpha} = \{c_j \,/\, c_j \in S(f) \text{ and } c_j \text{ contains the literal } x_i^{1-\alpha}, \ (j \in \{1, \ldots, m\})\}.$$

$d_nS(f)$ is the ordered set of ordered pairs of these subsets:

$$d_nS(f) = \{(fM_1^{\alpha}, fM_1^{1-\alpha}), (fM_2^{\alpha}, fM_2^{1-\alpha}), \ldots, (fM_n^{\alpha}, fM_n^{1-\alpha})\}.$$

We will use the notation $M^{\alpha}$ and $sM^{\alpha}$ also for the set $d_nS(f)$.

<u>*Lemma*</u> 7.1. For any function $f(x_1, \ldots, x_n)$, represented in $CNF$, the set $d_nS(f)$ is a special decomposition of the set $S(f)$.

<u>*Proof.*</u> Consider the conditions (2.1.1), (2.1.2) and (2.1.3).

$$(2.1.1) \quad \forall\, i \in \{1, \ldots, n\} \ \ (fM_i^{\alpha} \cap fM_i^{1-\alpha}) = \emptyset.$$

This is evident since none of the clauses contains the literals $x_i^{\alpha}$ and $x_i^{1-\alpha}$ simultaneously.

(2.1.2) $\forall\, i \in \{1,\dots,n\}$ $(fM_i^{\alpha} \neq \emptyset$ or $fM_i^{1-\alpha} \neq \emptyset)$

If $fM_i^{\alpha} = \emptyset$ and $fM_i^{1-\alpha} = \emptyset$ for some $i \in \{1,\dots,n\}$, then the literals $x_i^{\alpha}$ and $x_i^{1-\alpha}$ do not belong to any clause. And this contradicts the formation of the subsets $fM_i^{\alpha}$ and $fM_i^{1-\alpha}$.

(2.1.3) $\bigcup_{i=1}^{n}(fM_i^{\alpha} \cup fM_i^{1-\alpha}) = S(f)$,

If for some $j \in \{1,\dots,m\}$, $c_j \in \bigcup_{i=1}^{n}(fM_i^{\alpha} \cup fM_i^{1-\alpha})$, then for some $i \in \{1,\dots,n\}$, $c_j \in fM_i^{\alpha}$ or $c_j \in fM_i^{1-\alpha}$, so, $c_j \in S(f)$.

If $c_j \in S(f)$, then $c_j$ contains some literals. So, for some variable $x_i$ either $x_i^{\alpha}$ is found in the clause $c_j$, or $x_i^{1-\alpha}$ is found in the clause $c_j$. Thus,

$$c_j \in \bigcup_{i=1}^{n}(fM_i^{\alpha} \cup fM_i^{1-\alpha}).$$

Therefore, for any function $f(x_1,\dots,x_n)$, represented as $CNF$, the set $d_nS(f)$ is a special decomposition of the set $S(f)$. ∇

If under the special decomposition $d_nS(f)$, there exists a special covering for the set $S(f)$, then we will denote such a covering by

$$c_nS(f) = \{fM_1^{\alpha_1}, fM_2^{\alpha_2},\dots, fM_n^{\alpha_n}\}.$$

*Theorem* 7.2. For any function $f(x_1,\dots,x_n)$ represented in $CNF$, the following is true:

There are $\sigma_1,\dots,\sigma_n$ ($\sigma_i \in \{0,1\}$) such that

$$f(\sigma_1,\dots,\sigma_n) = 1$$

if and only if there is a special covering for the set $S(f)$ under the decomposition $d_nS(f)$.

*Proof*. Let $f(\sigma_1,\dots,\sigma_n) = 1$ for some $\sigma_1,\dots,\sigma_n$, where $\sigma_i \in \{0,1\}$.

We will show that then the set

$$c_nS(f) = \{fM_1^{\sigma_1}, fM_2^{\sigma_2},\dots, fM_n^{\sigma_n}\}$$

will be a special covering for the set $S(f)$ under the special decomposition $d_nS(f)$. To show this, we prove that

$$\bigcup_{i=1}^{n} fM_i^{\sigma_i} = S(f).$$

It is enough to show that each clause belongs to some subset, included in the set $c_nS(f)$.

Suppose that there is a clause $c_j \in S(f)$ that does not belong to any of the subset included in $c_nS(f)$. It means that none of the literals $x_1^{\sigma_1}, x_2^{\sigma_2},\dots, x_n^{\sigma_n}$ is found in the clause $c_j$.

Therefore, $c_j$ is the disjunction of some literals of the form $x_i^{1-\sigma_i}$.

Since $(\sigma_i^{1-\sigma_i} = 0)$, for any $i \in \{1,\dots,n\}$, then for given values of variables, $c_j = 0$. This contradicts the assumption that

$$f(\sigma_1, \sigma_2,\dots,\sigma_n) = 1.$$

So, each clause is included in some subset included in the set $c_nS(f)$.

Let for some $\alpha_1, \alpha_2,\dots,\alpha_n \in \{0,1\}$ the set

$$c_nS(f) = \{fM_1^{\alpha_1}, fM_2^{\alpha_2},\dots, fM_n^{\alpha_n}\}$$

is a special covering for the set $S(f)$ under the decomposition $d_nS(f)$.

By definition, the subset $fM_i^{\alpha_i}$ contains clauses containing the literal $x_i^{\alpha_i}$.

Therefore, if $x_i^{\alpha_i} = 1$, then the value of all clauses included in the set $fM_i^{\alpha_i}$ is equal 1, that is for any

$$i \in \{1, \ldots, n\} \text{ and } j \in \{1, \ldots, n\},$$

the following holds:

$$(x_i^{\alpha_i} = 1) \; \& \; (c_j \in fM_i^{\alpha_i})] \Rightarrow ( c_j = 1).$$

Obviously, if $\sigma_1=\alpha_1, \sigma_2=\alpha_2, \ldots, \sigma_n=\alpha_n$, then $f(\sigma_1, \ldots, \sigma_n) = 1$. ∇

Now, based on some special decomposition $d_nS$, we form a Boolean function, which will be represented in conjunctive normal form. Thus, each special decomposition will generate certain Boolean function. This function will be denoted as

$$h^I(x_1, \ldots, x_n),$$

where $x_1, \ldots, x_n$ are Boolean variables. However, if this does not lead to ambiguity, we often skip the superscript.

To form the function $h^I(x_1, \ldots, x_n)$, for each element $e_i \in S$ we form the set of literals denoted by $l(e_i)$ as follows:

if $e_i \in M_j^{\alpha_j}$, then we form the literal $x_j^{\alpha_j}$ and add it to the set $l(e_i)$.

It is easy to see, that as a result of the formation of literals $x_j^{\alpha_j}$, the number of variables will be equal to $n$.

In fact, for each element $e_i \in \{e_1, e_2, \ldots, e_m\}$ we will have:

$$l(e_i) = \{x_j^{\alpha_j} \; / \; e_i \in M_j^{\alpha_j}, \;\; j \in \{1, \ldots, n\}, \; \alpha_j \in \{0,1\} \}.$$

Let $c_i$ be the clause formed by the literals of the set $l(e_i)$. Obviously, the number of these clauses will be equal to $m$.

Then, we form the function $h^I$ as follows:

$$h^I(x_1, \ldots, x_n) = \wedge_{i=1}^{m} c_i.$$

It is easy to see that the generated function $h^I(x_1, \ldots, x_n)$ is a Boolean function in conjunctive normal form. It is also obvious that particular function in $CNF$ will correspond to any special decomposition.

*Theorem* 7.3. If the set

$$d_nS = \{(M_1^{\alpha}, \; M_1^{1-\alpha}), \ldots, (M_i^{\alpha}, \; M_i^{1-\alpha}), \ldots, (M_n^{\alpha}, \; M_n^{1-\alpha})\}$$

is a special decomposition of the set $S$, and $h(x_1, \ldots, x_n)$ is the function generated by this decomposition, then:

There exists a special covering for the set $S$ under the decomposition $d_nS$, if and only if there exists a tuple $(\sigma_1, \ldots, \sigma_n)$, $(\sigma_j \in \{0,1\})$ such that

$$h(\sigma_1, \ldots, \sigma_n) = 1.$$

*Proof.* Suppose that for some $\alpha_1, \alpha_2, \ldots, \alpha_n$ $(\alpha_j \in \{0,1\})$, the set

$$c_nS = \{M_1^{\alpha_1}, \; M_2^{\alpha_1}, \ldots, M_n^{\alpha_n}\}$$

is the special covering for the set $S$.

This means that for each $e_i \in S$ there exists a subset $M_j^{\alpha_j} \in c_nS$ such that $e_i \in M_j^{\alpha_j}$. So, by definition, the literal $x_j^{\alpha_j}$ occurs in the clause $c_i$. That is,

$\forall\ i \in \{1,\ldots,m\}$, if $e_i \in M_j^{\alpha_j}$, then the literal $x_j^{\alpha_j}$ is found in the clause $c_i$.

It is easy to notice, that if $\sigma_1 = \alpha_1, \ldots, \sigma_n = \alpha_n$, then $h(\sigma_1, \ldots, \sigma_n) =1$.

Suppose now, that $h(\sigma_1, \ldots, \sigma_n) =1$ for some $\sigma_1, \ldots, \sigma_n$, $(\sigma_j \in \{0,1\})$.

According to Theorem 7.2, the set

$$c_nS(h)= \{hM_1^{\sigma_1}, hM_2^{\sigma_2}, \ldots, hM_n^{\sigma_n}\}$$

is a special covering for the set $S(h)$ under the decomposition $d_nS(h)$.

Let us prove that then the set

$$c_nS = \{M_1^{\sigma_1}, M_2^{\sigma_2}, \ldots, M_n^{\sigma_n}\}$$

will be a special covering for the set $S$.

Since the set $c_nS(h)$ is a special covering for the set $S(h)$, for every clause $c_i$ there exists a subset $hM_j^{\sigma_j} \in c_nS(h)$ such that $c_i \in hM_j^{\sigma_j}$. This means that the clause $c_i$ contains the literal $x_j^{\sigma_j}$, since by definition

$$hM_j^{\sigma_j} = \{c_k \ / \ c_k \in S(h) \text{ and } c_k \text{ contains } x_j^{\sigma_j},\ ( k \in \{1,\ldots,m\})\}.$$

On the other hand, by definition of the formation of clauses, the clause $c_i$ contains the literal $x_j^{\sigma_j}$ only if $e_i \in M_j^{\sigma_j}$.

Since each element $e_i \in S$ determines the composition of one clause, and each clause is defined by one element of the set $S$, then it is easy to prove that for any element $e_i \in S$ there exists a subset $M_j^{\sigma_j} \in c_nS$ such, that $e_i \in M_j^{\sigma_j}$.

Therefore, the set $c_nS = \{M_1^{\sigma_1}, M_2^{\sigma_2}, \ldots, M_n^{\sigma_n}\}$ is a special covering for the set $S$. $\nabla$

In fact, the Theorems 7.2 and 7.3 prove that:

-each Boolean function $f(x_1, \ldots, x_n)$ of $n$ variables represented in $CNF$ with $m$ clauses, generates a special decomposition $d_nS(f)$ of the set $S(f)$.

- each special decomposition of any set of $m$ elements and containing $n$ ordered pairs, generates a Boolean function of $n$ variables in $CNF$ with $m$ clauses. We denoted it as

$$h^I(x_1, \ldots, x_n).$$

- the Boolean satisfiability problem and the problem of existence of special covering of the set are equivalent in the sense that the decidability algorithm for any of these problems generates a decidability algorithm for another.

Moreover, as it is proved in [13], the mentioned problems are polynomially equivalent, each of them is polynomially reducible to the other.

Therefore, the problem of existence of special covering of the set under the special decomposition is an $NP$-complete problem.

## 8. Applications

Given a Boolean function $f(x_1, \ldots, x_n)$ of $n$ variables, which is represented in $CNF$ by $m$ clauses. Let

$$S(f) = \{c_1, c_2, \ldots, c_m\}$$

be the set of clauses of the function $f(x_1, \ldots, x_n)$, and

$$d_nS(f) = \{(fM_1^{\alpha}, fM_1^{1-\alpha}), \ldots, (fM_i^{\alpha}, fM_i^{1-\alpha}), \ldots, (fM_n^{\alpha}, fM_n^{1-\alpha})\}$$

be a special decomposition of the set $S(f)$.

According to Theorem 7.2, the function $f(x_1, \ldots, x_n)$ represented in $CNF$ is satisfiable if and only if there is a special covering for the set $S(f)$ under the decomposition $d_nS(f)$.

Hence, considering the set of clauses, $S(f)$, by Theorem 6.13, we find out if there exists a special covering for the set $S(f)$ under the special decomposition $d_nS(f)$. Then, using Theorem 7.2, we find out whether the function $f(x_1, \ldots, x_n)$ is satisfiable. Let's formulate this as a theorem.

*Theorem* 8.1. Given a Boolean function $f(x_1, \ldots, x_n)$ represented in $CNF$.

$S(f) = \{c_1, \ldots, c_m\}$ is the set of clauses of the function $f(x_1, \ldots, x_n)$.

The set of ordered pairs

$$d_nS(f) = \{(fM_1^{\alpha}, fM_1^{1-\alpha}), \ldots, (fM_i^{\alpha}, fM_i^{1-\alpha}), \ldots, (fM_n^{\alpha}, fM_n^{1-\alpha})\}$$

is a special decomposition of the set $S(f)$ such that

$$\{c_{q_1}, \ldots, c_{q_l}\} = S(f) \backslash M^{\alpha},$$

$G(c_{q_1}, \ldots, c_{q_l})$ is the pointed graph corresponding to the replaceability procedure associated with the set $\{c_{q_1}, \ldots, c_{q_l}\}$ under the decomposition $d_nS(f)$.

$ExG(e_{s_1}, \ldots, e_{s_p})$ is the final extension of the clean graph $CG(e_{q_1}, \ldots, e_{q_l})$.

Then, the function $f(x_1, \ldots, x_n)$ is satisfiable if and only if the set

$$\{c_{\mathrm{q}}, \ldots, c_{q_l}, c_{s_1}, \ldots, c_{s_p}\}$$

is stable with respect to both the cleaning procedure and the compatibility procedure in the graph $ExG(e_{s_1}, \ldots, e_{s_p})$.

*Proof.* The proof follows from Theorems 6.13 and 7.2. ∇

Let the function $f(x_1, \ldots, x_n)$ is represented in $CNF$, and the set $d_nS(f))$ is a special decomposition of the set $S(f)$.

For some $\{i_1, \ldots, i_k\} \subseteq \{1, \ldots, n\}$ we denote the following notation:

$$(i_1, \ldots, i_k)h^I(x_1, \ldots, x_n)$$

is the function generated by the special decomposition $(i_1, \ldots, i_k)I(d_nS(f))$.

We will use also the notation $h^I(x_1, \ldots, x_n)$ for this function if this does not lead to ambiguity.

*Theorem* 8.2. For any $\{i_1, \ldots, i_k\} \subseteq \{1, \ldots, n\}$, the function $f(x_1, \ldots, x_n)$ represented in $CNF$ is satisfiable if and only if the function $\{i_1, \ldots, i_k\}h^I(x_1, \ldots, x_n)$ is satisfiable.

*Proof.* Let the set of ordered pairs

$$d_nS(f) = \{(fM_1^{\alpha}, fM_1^{1-\alpha}), \ldots, (fM_i^{\alpha}, fM_i^{1-\alpha}), \ldots, (fM_n^{\alpha}, fM_n^{1-\alpha})\}$$

be a special decomposition of the set $S(f)$.

Consider the special decomposition $(i_1, \ldots, i_k)I(d_nS(f))$. By definition,

$(i_1, \ldots, i_k)I(d_nS(f)) = \{(fM_1^{\sigma_1}, fM_1^{1-\sigma_1}), \ldots, (fM_i^{\sigma_i}, fM_i^{1-i}), \ldots, (fM_n^{\sigma_n}, fM_n^{1-n})\}$, for

$$\sigma_i = \begin{cases} \alpha, & i \notin \{ i_1, \ldots, i_k\} \\ 1-\alpha, & i \in \{ i_1, \ldots, i_k\} \end{cases}$$

Suppose, that under the decomposition $d_nS(f)$, for some $i \in \{i_1, \ldots, i_k\}$, the ordered pair $(fM_i^{\alpha}, fM_i^{1-\alpha})$ consists of following components:

$$fM_i^{\alpha} = \{c_{l_1}^i, \ldots, c_{l_{p(i)}}^i\} \text{ and } fM_i^{1-\alpha} = \{c_{j_1}^i, \ldots, c_{j_{q(i)}}^i\},$$

where $\{c_{l_1}^i, \ldots, c_{l_{p(i)}}^i\} \subseteq S(f)$ and $\{c_{j_1}^i, \ldots, c_{j_{q(i)}}^i\} \subseteq S(f)$,

$p(i)$ and $q(i)$ are the numbers of clauses in the subsets $fM_i^{\alpha}$ and $fM_i^{1-\alpha}$ respectively.

This means that:

- the literal $x_i^{\alpha}$ is included in all clauses of the set $\{c_{l_1}^i, \ldots, c_{l_p}^i\}$,

- the literal $x_i^{1-\alpha}$ is included in all clauses of the set $\{c_{j_1}^i, \ldots, c_{j_q}^i\}$.

-the literals $x_i^{\alpha}$ and $x_i^{1-\alpha}$ are not included in any other clauses.

After permuting the components of the ordered pair $(fM_i^{\alpha}, fM_i^{1-\alpha})$, all elements of the set $\{c_{j_1}^i, \ldots, c_{j_{q(i)}}^i\}$ are moved to the $\alpha$-domain, and all elements of the set $\{c_{l_1}^i, \ldots, c_{l_{p(i)}}^i\}$ are moved to the (1-$\alpha$)-domain.

Since $h^I(x_1, \ldots, x_n)$ is generated by the special decomposition $(i_1, \ldots, i_k)I(d_nS(f))$, then it is obvious, that any clause of the function $h^I(x_1, \ldots, x_n)$ belonging to the set

$$S(f) \setminus \cup_{i \in \{i_1,\ldots,i_k\}}(\{c_{l_1}^i, \ldots, c_{l_{p(i)}}^i\} \cup \{c_{j_1}^i, \ldots, c_{j_{q(i)}}^i\})$$

coincides with the corresponding clause of the function $f(x_1, \ldots, x_n)$.

It is also clear, that for each $i \in \{i_1, \ldots, i_k\}$:

- in all clauses of the function $h^I(x_1, \ldots, x_n)$ belonging to the set $\{c_{j_1}^i, \ldots, c_{j_{q(i)}}^i\}$, the literal $x_i^{1-\alpha}$ is replaced by the literal $x_i^{\alpha}$,

- in all clauses of the function $h^I(x_1, \ldots, x_n)$ belonging to the set $\{c_{l_1}^i, \ldots, c_{l_{p(i)}}^i\}$, the literal $x_i^{\alpha}$ is replaced by the literal $x_i^{1-\alpha}$,

Suppose that the function $f(x_1, \ldots, x_n)$ is satisfiable, that is, there is a tuple $(\sigma_1, \ldots, \sigma_n)$, $(\sigma_j \in \{0,1\}$ such that $f(\sigma_1, \ldots, \sigma_n) = 1$.

It is easy to see that the function $(i_1, \ldots, i_k)h^I(x_1, \ldots, x_n)$ will be satisfiable by the tuple $(\beta_1, \ldots, \beta_n)$ if we define this tuple as follows:

$$\beta_i = \begin{cases} \sigma_i, & i \notin \{ i_1, \ldots, i_k\} \\ 1-\sigma_i, & i \in \{ i_1, \ldots, i_k\} \end{cases}.$$

That is, $(i_1, \ldots, i_k)h^I(\beta_1, \ldots, \beta_n) = 1$.

Obviously, the opposite is also true. if $h^I(\beta_1, \ldots, \beta_n) = 1$ for some $\beta_1, \ldots, \beta_n$, then there is a tuple $(\sigma_1, \ldots, \sigma_n)$ such that $f(\sigma_1, \ldots, \sigma_n) = 1$. ∇

It is easy to notice that the function $(i_1, \ldots, i_k)h^I(x_1, \ldots, x_n)$ is obtained by replacing the literals $x_{i_1}^{\alpha}, \ldots, x_{i_k}^{\alpha}$ with the literals $x_{i_1}^{1-\alpha}, \ldots, x_{i_k}^{1-\alpha}$, respectively, and the literal $x_{i_1}^{1-\alpha}, \ldots, x_{i_k}^{1-\alpha}$ with the literal $x_{i_1}^{\alpha}, \ldots, x_{i_k}^{\alpha}$ in all clauses of the function $f(x_1, \ldots, x_n)$.

*We will say that the function* $(i_1, \ldots, i_k)h^I(x_1, \ldots, x_n)$ *is obtained by inverting the literals of the variables* $x_{i_1}, \ldots, x_{i_k}$ *in the function* $f(x_1, \ldots, x_n)$.

Based on Theorem 8.1, we formulate another necessary and sufficient conditions for Boolean satisfiability.

*Definition* 8.3. A conjunctive normal form of a Boolean function will be called a *Proportional Conjunctive Normal Form,* if each clause contains at least one negative literal or if each clause contains at least one positive literal.

*Theorem* 8.4. A function $f(x_1, \ldots, x_n)$ represented in $CNF$ is satisfiable if and only if it can be transformed into a function in Proportional $CNF$ using literal inversions.

*Proof.* Assume that the function $f(x_1, \ldots, x_n)$ represented in $CNF$ is satisfiable, and the set $d_nS(f)$ is a special decomposition corresponding to this function.

This means that there are $\sigma_1, \ldots, \sigma_n$ ($\sigma_j \in \{0,1\}$) such that

$$f(\sigma_1, \ldots, \sigma_n) = 1.$$

Hence, according to Theorem 7.2, there is a special covering for the set $S(f)$ under the decomposition $d_nS(f)$. Let it be the set

$$c_nS = \{fM_1^{\sigma_1}, fM_2^{\sigma_2}, \ldots, fM_n^{\sigma_n}\}.$$

Obviously, if the set $c_nS$ is an $M^{\alpha}$-covering for the set $S(f)$, then all clauses are found in the $\alpha$-domain of the corresponding decomposition, so all they contain a negative literal, if $\alpha$=0, or all they contain a positive literal, if $\alpha$=1.

Assume that $c_{q_1}, \ldots, c_{q_l}$ are clauses of the function $f$ such that for some $\alpha \in \{0,1\}$,

$$\{c_{q_1}, \ldots, c_{q_l}\} = S(f) \setminus M^{\alpha}.$$

Since there is a special covering for the set $S(f)$ then according to Theorem 3.5, the set $\{c_{q_1}, \ldots, c_{q_l}\}$ is reachable. Suppose that

$$\{(fM_{i_1}^{\alpha}, fM_{i_1}^{1-\alpha}), \ldots, (fM_{i_k}^{\alpha}, fM_{i_k}^{1-\alpha})\}$$

is the set of replacement steps of the decomposition $d_nS(f)$ that ensure the reachability of all elements $\{c_{q_1}, \ldots, c_{q_l}\}$.

After the permutation the components of all these ordered pairs we obtain the special decomposition $(i_1, \ldots, i_k)I(d_nS(f))$. Obviously, as a result, all clauses of the set $S(f)$ will be in the $\alpha$-domain of the resulting decomposition.

Consider the function $(i_1, \ldots, i_k)h^I(x_1, \ldots, x_n)$ generated by the special decomposition $(i_1, \ldots, i_k)I(d_nS(f))$.

Since all clauses of this function are included in the $\alpha$-domain of the decomposition

$$(i_1, \ldots, i_k)I(d_nS(f)),$$

then all they contain a negative literal, if $\alpha$=0, or all they contain a positive literal, if $\alpha$=1.

This means that, the function

$$(i_1, \ldots, i_k)h^I(x_1, \ldots, x_n)$$

is represented in proportional $CNF$.

At the other hand, obviously, the function $(i_1, \ldots, i_k)h^I(x_1, \ldots, x_n)$ is obtained by inverting the literals of the variable $x_{i_1}, \ldots, x_{i_k}$ in the function $f(x_1, \ldots, x_n)$.

So, if $f(x_1, \ldots, x_n)$ is satisfiable, then it can be transformed into a function represented in Proportional $CNF$ using literal inversions.

It is easy to see, that the opposite also is true:

Suppose that the function is transformed into a function represented in Proportional $CNF$ using literal inversions. Let it be the function $h^I(x_1, \ldots, x_n)$.

Since any function in proportional $CNF$ is satisfiable, then by the Theorem 8.2, the function $f(x_1, \ldots, x_n)$ is satisfiable. ∇

8.4 *Finding the Maximum Number of Satisfiable Clauses of a Boolean Function*

Suppose that the function $f(x_1, \ldots, x_n)$, represented in $CNF$ is not satisfiable and we need to find the maximum number of satisfiable clauses of this function, provided that this function is not satisfiable.

To do this, we consider the special decomposition $d_nS(f)$ of the set $S(f)$ and apply the procedures used in search for a special covering for the set $S(f)$ to the decomposition $d_nS(f)$.

Assume that we have obtained a special decomposition and elements(clauses) that cannot be added to the $\alpha$-domain using the replaceability procedure.

In this case the description from point 6.14 is applied to the resulting decomposition. As a result, we will get maximum number of clauses that can simultaneously be in the $\alpha$-domain.

The following theorem will show that if we have the maximum number of clauses in the a-domain, then this will be the maximum number of satisfiable clauses.

*Theorem* 8.5 Let $f(x_1, \ldots, x_n)$ is a Boolean function represented in $CNF$, and $S(f)$ be a set of clauses of this function. Then the following holds:

For any $I$-transformation of the decomposition $d_nS(f)$, there are tuples $\alpha_1, \ldots, \alpha_n$ and $\beta_1, \ldots, \beta_n$, where $\alpha_i$, $\beta_j \in \{0,1\}$, such that:

All clauses in the $\alpha$-domain are satisfiable when giving the values $\alpha_1, \ldots, \alpha_n$ to the variables $x_1, \ldots, x_n$,

All clauses in the $(1\text{-}\alpha)$-domain are satisfiable when giving the values $\beta_1, \ldots, \beta_n$ to the variables $x_1, \ldots, x_n$.

*Proof.* Let for some $\sigma_1, \ldots, \sigma_n$, ($\sigma_j \in \{0,1\}$), the set

$$(r_1, \ldots, r_p)I(d_nS(f))=\{(fM_1^{\sigma_1}, fM_1^{1-\sigma_1}), \ldots, (fM_n^{\sigma_n}, fM_n^{1-\sigma_n})$$

be an $I$-transformation of the decomposition $d_nS(f)$. This means that

$$\sigma_i = \begin{cases} \alpha, & i \notin \{ r_1, \ldots, r_p\} \\ 1-\alpha, & i \in \{ r_1, \ldots, r_p\} \end{cases}$$

Consider an arbitrary subset in the $\alpha$-domain of this decomposition

$$fM_k^{\sigma_k} \in \{fM_1^{\sigma_1}, \ldots, fM_n^{\sigma_n}\}$$

and the clauses included in it. By definition

$$fM_k^{\sigma_k}= \{c_j \,/\, c_j \in S(f) \text{ and } c_j \text{ contains the literal } x_k^{\sigma_k}, \ ( j \in \{1, \ldots, m\})\}.$$

Obviously, if $\alpha_k= \sigma_k$ then all clauses in the subset $fM_k^{\sigma_k}$ will take the value 1. So, if

$$\alpha_1= \sigma_1, \ldots, \alpha_n= \sigma_n,$$

then all clauses included in the $\alpha$-domain of this decomposition will be satisfiable.

Using the definition for $\sigma_i$, we get

$$1\text{-}\sigma_i = \begin{cases} \alpha, & i \notin \{ r_1, \ldots, r_p\} \\ 1-\alpha, & i \in \{ r_1, \ldots, r_p\} \end{cases}.$$

By definition of the subsets in the (1-$\alpha$)-domain of the same decomposition,

$$fM_k^{1-\sigma_k}= \{c_j \,/\, c_j \in S(f) \text{ and } c_j \text{ contains the literal } x_k^{1-\sigma_k}, \ ( j \in \{1, \ldots, m\})\}.$$

So, by the same reasoning as for the $\alpha$-domain, taking

$$(\beta_1=1\text{-}\sigma_1), \ldots, (\beta_n=1\text{-}\sigma_n),$$

we can assert that all clauses included in the (1-$\alpha$)-domain of the same decomposition will be satisfiable. ∇

*Corollary* 8.5.1 If $f(x_1, \ldots, x_n)$ is a Boolean function represented in $CNF$, and $S(f)$ is a set of clauses of this function, then the following is true:

If the function $f(x_1, \ldots, x_n)$ is not satisfiable, then the set $S(f)$ can be divided into two nonempty subsets so that all the clauses included in any of them are satisfiable.

*Proof.* Consider any $I$-transformation, $I(d_nS(f))$, of the special decomposition $d_nS(f)$. By Theorem 8.5 there are tuples $\alpha_1, \ldots, \alpha_n$ and $\beta_1, \ldots, \beta_n$, where $\alpha_i$, $\beta_j \in \{0,1\}$, such that:

- the clauses included in the $\alpha$-domain of this decomposition are satisfiable when giving the values $\alpha_1, \ldots, \alpha_n$ to the variables $x_1, \ldots, x_n$, respectively,

- the clauses included in the (1-$\alpha$)-domain of this decomposition are satisfiable when giving the values $\beta_1, \ldots, \beta_n$ to the variables $x_1, \ldots, x_n$, respectively.

Let $S_1(f)$ be an arbitrary chosen nonempty subset of the elements included in the $\alpha$-domain and $S_2(f)$ an arbitrary chosen nonempty subset of the elements included in the (1-$\alpha$)-domain such that

$$S_1(f) \cup S_2(f) = S(f).$$

It is easy to see that $S_1(f)$ and $S_2(f)$ satisfy the required conditions. ∇

9. *The General Algorithm for Determining the Satisfiability of a Boolean Function* $f(x_1, \ldots, x_n)$ *in* $CNF$.

1. Compose the special decomposition, $d_nS(f)$ corresponding to $f(x_1, \ldots, x_n)$;
2. Find the set $X$ of elements not included in the $\alpha$-domain;

   // $X = \{c_{q_1}, \ldots, c_{q_l}\} = S(f) \backslash M^{\alpha}$//

3. **If** $X = \emptyset$:
4.     **Output** ("$\alpha$-domain of the given decomposition is a special covering for the set $S$");
5.     **Stop**;
6. **Else**:
7.     Construct the pointing graph $G$ associated with elements $\{c_{q_1}, \ldots, c_{q_l}\}$;
8.     Search for useless vertices in the graph;
9.     **While** a useless vertex is found:
10.         Apply the *Removal Procedure* to this vertex;
11.         **If** this is not a removable vertex:
12.             **Output** "There is no special covering";
13.             **Stop**;
14.         **Else:**
15.             Repeat searching for another useless vertex;
16.         **EndIf**;
17.     **EndWhile;**
18.     IncompSetIndicator := 0;
19.     Search for incompatible set in the pointing graph;
20.     **While** there is an incompatible set with respect to some element $e \in S \setminus M^{\alpha}$:
21.         IncompatibleSetIndicator := 1;
22.         Apply the *Compatibility Procedure* to the found set;
23.         **If** incompatibility of the set under consideration is eliminated:
24.             Repeat searching for incompatible set with respect to another element $e$;
25.         **Else:**
26.             **If** $e$ is not included in (1-$\alpha$)-components of some ordered pairs that are not included in the replaceability procedure:
27.                 **Output** ("There is no special covering");
28.                 **Stop;**
29.             **Else:**
30.                 Add $e$ to the set $Y$;     // $Y$ is the set of elements by which the considered graph will be extended //
31.                 Repeat searching for incompatible set with respect to another element $e$;

```
32.           EndIf;
33.        EndIf;
34.     EndWhile;

35.     If IncompatibleSetIndicator = 0:
36.        Output "α-domain is a special covering"
37.        Stop;
38.     Else:

39.        If  Y = ∅:
40.            Output ("α-domain is a special covering")
41.            Stop;
42.        Else:
43.            Add to the graph under construction new main vertices associated
                                      with the elements included in Y;
44.            Go to 7, the Constructing Procedure for the extended graph;
45.        EndIf;

46.    EndIf;
47.  EndIf;
```

Descriptions of the individual procedures included in this algorithm and estimates of their complexity, as well as the estimate of the complexity of the general algorithm, will be presented in the next article.

We will prove that the complexity of all these procedures is polynomial.